\documentclass[10pt,journal]{IEEEtran} 
\IEEEoverridecommandlockouts

\usepackage{amsfonts, amsmath,amssymb,amsthm}
\usepackage{algorithm, algorithmic}
\usepackage{array}
\usepackage{cases}
\usepackage{changepage}
\usepackage[compress]{cite}
\usepackage{epstopdf}
\usepackage[english]{babel}
\usepackage{etoolbox}
\usepackage{graphicx}
\usepackage{multirow}
\usepackage{physics} 	
\usepackage{subfigure}
\usepackage{setspace}
\usepackage{url}  
\usepackage[hyphenbreaks]{breakurl} 
\usepackage{lipsum}
\usepackage{nicefrac}	
\usepackage[table,xcdraw]{xcolor}	
\usepackage{textcomp}
\usepackage{makecell}
\usepackage[font=small]{caption} 
\usepackage{mathrsfs}
\usepackage{indentfirst} 

\def\ps@headings{%
\def\@oddhead{\mbox{}\scriptsize\rightmark \hfil \thepage}%
\def\@evenhead{\scriptsize\thepage \hfil \leftmark\mbox{}}%
\def\@oddfoot{}%
\def\@evenfoot{}}
\makeatother
\makeatletter
\patchcmd{\@thm}
  {\trivlist}
  {\list{}{\leftmargin=\thm@leftmargin\rightmargin=\thm@rightmargin}}
  {}{}
\patchcmd{\@endtheorem}
  {\endtrivlist}
  {\endlist}
  {}{}
\newlength{\thm@leftmargin}
\newlength{\thm@rightmargin}

\newcommand{\xnewtheorem}[3]{%
  \newenvironment{#3}
    {\thm@leftmargin=#1\relax\thm@rightmargin=#2\relax\begin{#3INNER}}
    {\end{#3INNER}}%
  \newtheorem{#3INNER}%
}

\makeatother

\newtheoremstyle{indentedupright}{3pt}{3pt}{} {}{\bfseries}{.}{.5em}{} 
\newtheoremstyle{indenteditalic}{3pt}{3pt}{\itshape} {}{\bfseries}{.}{.5em}{} 

\theoremstyle{indenteditalic}
\xnewtheorem{0pt}{0pt}{oldpro}{Conventional Property}
\xnewtheorem{0pt}{0pt}{newpro}{New Property}
\xnewtheorem{0pt}{0pt}{obser}{Observation}
\xnewtheorem{0pt}{0pt}{pro}{Property}
\xnewtheorem{0pt}{0pt}{lem}{Lemma}
\xnewtheorem{0pt}{0pt}{corl}{Corollary}
\xnewtheorem{0pt}{0pt}{addcon}{Additional Constraint}

\newcommand{\romu}[1]{\uppercase\expandafter{\romannumeral #1\relax}} 
\newcommand{\roml}[1]{\lowercase\expandafter{\romannumeral #1\relax}}    

\renewcommand{\baselinestretch}{0.96}

\begin{document}
\title{\LARGE Enhancing In-Situ Structural Health Monitoring through RF Energy-Powered Sensor Nodes and Mobile Platform}
\author{\IEEEauthorblockN{Yu Luo\IEEEauthorrefmark{1}, Lina Pu\IEEEauthorrefmark{2}, Jun Wang\IEEEauthorrefmark{3}, Isaac Howard\IEEEauthorrefmark{3}}\\
\IEEEauthorblockA{\IEEEauthorrefmark{1} Department of Electrical and Computer Engineering, Mississippi State University, Mississippi State, MS, 39762\\
\IEEEauthorrefmark{2}Department of Computer Science, University of Alabama, Tuscaloosa, AL 35487\\
\IEEEauthorrefmark{3}Richard A. Rula School of Civil \& Environmental Engineering, Mississippi State University,  MS, 39762\\
Email: yu.luo@ece.msstate.edu, lina.pu@ua.edu, jwang@cee.msstate.edu, ilhoward@cee.msstate.edu}
}

\maketitle

\begin{abstract}
\label{:Abstract}
This research contributes to long-term structural health monitoring (SHM) by exploring radio frequency energy-powered sensor nodes (RF-SNs) embedded in concrete. Unlike traditional in-situ monitoring systems relying on batteries or wire-connected power sources, the RF-SN captures radio energy from a mobile radio transmitter for sensing and communication. This offers a cost-effective solution for consistent in-situ perception. To optimize the system performance across various situations, we've explored both active and passive communication methods. For the active RF-SN, we implement a specialized control circuit enabling the node to transmit data through ZigBee protocol at low incident power. For the passive RF-SN, radio energy is not only for power but also as a carrier signal, with data conveyed by modulating the amplitude of the backscattered radio wave. To address the challenge of significant attenuation of the backscattering signal in concrete, we utilize a square chirp-based modulation scheme for passive communication. This scheme allows the receiver to successfully decode the data even under a negative signal-to-noise ratio (SNR) condition. The experimental results indicate that an active RF-SN embedded in concrete at a depth of 13.5\,cm can be effectively powered by a 915\,MHz mobile radio transmitter with an effective isotropic radiated power (EIRP) of 32.5\,dBm. This setup allows the RF-SN to send over 1 kilobyte of data within 10 seconds, with an additional 1.7 kilobytes every 1.6 seconds of extra charging. For the passive RF-SN buried at the same depth, continuous data transmission at a rate of 224\,bps with a 3\% bit error rate (BER) is achieved when the EIRP of the transmitter is 23.6\,dBm.

\end{abstract}


\section{Introduction}
\label{sec:Introduction}

Concrete is one of the most popular building materials in modern cities and suburban areas. On June 24, 2021, a 12-story reinforced concrete condo partially collapsed near Miami Beach left 98 dead~\cite{the2021what}. According to the investigation, the collapse began when the ground-floor parking area and pool deck have been damaged and neglected over the past 40 years~\cite{jade2021one, lu2021preliminary}. After this accident, ever-increasing attention is being paid to developing efficient methods for the continuous monitoring of building health.

Undoubtedly, structural health monitoring (SHM) plays a pivotal role in modeling and forecasting the durability of concrete~\cite{alexander2019durability}, a crucial step in averting unexpected building failures. The SHM procedure often requires focused assessments in specific areas. It can be laborious, necessitating frequent expert evaluations of the target structure. Further, visual inspections can prove challenging, particularly when trying to discern the internal condition of the concrete or when load-bearing structures are obscured by decorative elements.

Over the past few years, there has been a marked upswing in the research of sensor technology in SHM, enhancing both the accuracy and efficiency of detection. By choosing the appropriate sensors, various parameters such as strain, temperature, pressure, and moisture of concrete can be captured in real-time~\cite{berrocal2021crack, chang2012implementing}. These critical measurements provide indispensable insights for assessing structural health. Now, we are just one final piece away from completing the entire puzzle of persistent SHM --- the power supply.

In general, SHM systems are either powered by batteries or wired to an external power source. However, the use of batteries is typically limited to short-term monitoring, as battery replacement becomes impossible once the sensors are embedded in concrete. However, wiring to an external power source considerably raises the system's cost. As referenced in \cite{lynch2000development}, the setup of such a monitoring system may account for up to 25\% of the overall system expense, with over 75\% of the installation time dedicated specifically to the wiring process. Contemporary SHM system is seeking innovative solutions to liberate the embedded sensor nodes from the power supply constraint.

In response to the above requirements, our research explores the potential for harnessing radio frequency (RF) energy to power sensors embedded within concrete. In this approach, the sensor nodes are free from batteries and wiring to external power sources. They capture ultra-high frequency (UHF) radio energy radiated by a mobile RF transmitter for both sensing and communication. With the energy harvesting capability, RF energy-powered sensor nodes (RF-SNs) can function semi-perpetually within the concrete. This provides a cost-effective solution for future SHM systems.

The practical application of RF-SNs faces two significant hurdles. The first issue is attributed to the severe attenuation of radio energy in concrete coupled with a significant reflection loss of radio waves at the air-concrete interface. This leads to a thin RF energy that can be harvested by an RF-SN, potentially even lower than the circuit's leakage power. As a result, the sensor node cannot be started successfully, irrespective of how long it is charged. The second issue emerges from the high power consumption of commercial wireless modules, which can reach tens of milliwatts or beyond~\cite{microchip2020rn, microchip2016atmel}. This power demand necessitates extended charging durations for data transmission, thereby substantially diminishing its efficiency in real-world applications.

In order to tackle the aforementioned challenges,  we develop both active and passive RF-SNs to enhance system performance across a variety of scenarios. Specifically, in instances where RF-SNs are integrated into thin concrete structures (with a thickness of less than 20\,cm), the received energy is relatively strong. Under these circumstances, an active wireless module is triggered for communication using the ZigBee protocol. For the reliable startup of the active RF-SN to facilitate high-speed data transmission, we design a specialized control circuit that helps minimize power leakage and ensures the effective operation of sensor nodes in low-energy environments.

On the other hand, when RF-SNs are embedded in thick concrete structures (with a thickness exceeding 20\,cm), the incoming power intensity is weakened. Consequently, the harvested energy might be insufficient to power the active communication circuit. In this case, the node switches to a passive communication mode. This approach, akin to RF identification (RFID), not only harvests incoming radio waves as energy but also leverages them as carrier signals. Therefore, the RF-SN does not need to actively send out data. Rather, it modulates the amplitude of the reflected wave by modifying the impedance of the matching circuit behind the receiving antenna. As a result, the passive RF-SN could markedly conserve energy, making it ideal for low-speed wireless communication.

The strength of backscattering signals from passive RF-SNs is usually weak due to the significant propagation loss of UHF radio waves within concrete structures.  In order to reduce the bit error rate (BER), we employ a square chirp-based modulation scheme rather than the on-off keying (OOK) modulation typically used in commercial RFID for passive data transmission. This technique can greatly expand the communication bandwidth to facilitate the receiver in decoding data at a low signal-to-noise ratio (SNR).

We carried out comprehensive experiments to assess the functionality of the proposed system. The results show that when the effective isotropic radiated power (EIRP) of the external RF transmitter is 1.78\,W (equivalent to 32.5\,dBm) and the radio frequency is 915\,MHz, the RF-SN embedded in concrete at a depth of 13.5\,cm can receive radio energy with an intensity of 2\,dBm. In such conditions, the RF-SN can be successfully activated after a 10-second period of wireless charging,  and then transmits more than 1\,KB of data using a 3.5\,dBm transmission power. Subsequently, it is capable of sending an extra 1.7\,KB of data for each additional 1.6 seconds of charging.

Upon transitioning to the passive mode for backscattering communication, the RF-SN turns off the active wireless module and the power consumption reduces to 8.1\,$\mu$W. This adaptation facilitates successful data transmission under low incident power circumstances, considerably decreasing the required transmission power from the mobile radio transmitter. Consequently, the nodes can be incorporated into thicker concrete structures, while adhering to the maximum EIRP limit of 4\,W prescribed by the Federal Communications Commission (FCC) for industrial, scientific, and medical (ISM) equipment~\cite{joe2007federal}. According to our experimental results, the RF transmitter only requires an EIRP of 0.23\,W (equivalent to 23.6\,dBm) to maintain continuous sensing and data transmission of passive RF-SNs embedded in concrete at a depth of 13.5\,cm. Under these conditions, the bandwidth of the backscattering signal is 4.1\,kHz, with a data rate and BER of 224\,bps and 3\%, respectively.

To summarize, the major contributions of this work lie in the following three aspects:
\vspace{0.1cm}
\begin{adjustwidth}{-0.77cm}{0cm}
\begin{description}
\setlength{\labelsep}{-0.95em}
\itemsep 0.07cm
	\item[a)] We propose a promising solution for long-term in-situ SHM. It allows sensor nodes embedded in concrete to harvest RF energy from mobile radio sources. This eliminates the need for complex wire connections or periodical battery replacements, enabling sustainable sensing and wireless communication.
	\item[b)] Two different types of RF-SNs are developed to optimize system performance in various application scenarios. The active RF-SN, designed for embedding in thin concrete structures, ensures reliable data transmission at high data rates. On the other hand, the passive RF-SN, integrated into thick concrete structures, is optimized to minimize system power consumption.
	\item[c)] Both active and passive RF-SNs are implemented using commercial components. We have conducted extensive experiments under varying conditions to evaluate the system's performance. The results show that the proposed RF-SNs can operate effectively in thick concrete. This makes them an appealing solution for long-term, cost-effective, and automated in-situ SHM.
\end{description}
\end{adjustwidth}
\vspace{0.1cm}

The remainder of this paper is structured as follows. Section~\!\ref{sec:RelMov} presents the related work and motivations of our research. Section~\!\ref{subsec:App} gives examples of typical application scenarios for RF-SN. The impact of concrete material on the reflection properties of the antenna and the propagation loss of RF energy are examined in Section~\!\ref{sec:RfEng}. The architectures of the active and passive RF-SNs are elaborated in Section~\!\ref{sec:ActArch} and Section~\!\ref{sec:PassArch}, respectively. The experimental results are presented in Section~\!\ref{sec:ExpRes}. Finally, we conclude our work in Section~\ref{sec:Con}.

\section{Related Work and Motivation}
\label{sec:RelMov}

\subsection{Related Work}
\label{subsec:related}
To develop efficient RF-SNs for SHM, it is crucial to have a comprehensive understanding of how radio waves propagate within concrete materials. Early studies presented in \cite{taylor1997measurement} discovered that RF attenuation through concrete walls intensifies with increasing radio frequency. Afterward, the simulation results reported in \cite{jiang2011optimum} indicate that the transmission loss of RF waves due to reflection at the boundary of air and concrete is directly proportional to the moisture content in the concrete and inversely proportional to the RF frequency. Moreover, the propagation loss of RF waves within the concrete amplifies with increasing frequency and concrete moisture. Through computational analysis, it has been determined that the most effective frequency range for delivering RF energy into concrete is between 20\,MHz and 80\,MHz.

The research in \cite{dalke2000effects} delves into the impact of embedded rebar within concrete on the quality of RF signals. The findings underline that the strength of the received signal is greatly affected by the relative positioning of the rebar mesh and RF receiver. Notably, if the rebar mesh is located above the receiver, it functions as an inductive barrier for radio waves with wavelengths exceeding the size of the grid, consequently blocking the signals. Alternatively, when the rebar mesh is situated beneath the receiver, it operates as a reflector, enhancing the received signal at low frequency.  The research in \cite{asp2019impact} provides a more in-depth exploration of how moisture content within concrete impacts RF propagation. The experimental measurements reveal that at an RF frequency of 1 GHz, the propagation attenuation is roughly 1\,dB per 1.6\,cm in dry concrete. However, this attenuation increases to 9\,dB when dealing with wet concrete. Furthermore, if the RF frequency rises to 5\,GHz, the propagation attenuation in a 1.6\,cm of dry concrete is approximately 3\,dB. Yet, the attenuation escalates to 35\,dB when the radio wave with the same frequency passes through wet concrete.

In \cite{sidibe2021energy}, researchers developed a compact 3D T-match dipole antenna and an associated rectifier to effectively harvest RF energy at 818\,MHz in concrete. The experimental outcomes show that when the rectenna is buried 14\,cm deep in reinforced concrete and the EIRP of the external RF transmitter is set at 32\,dBm, the rectenna is capable of supplying 213\,$\mu$W of power and 1.46\,V of voltage to a load with a 10\,k$\Omega$ internal resistance.

The methods of encapsulating wireless sensor nodes for in-situ SHM are explored in references \cite{chang2012implementing} and \cite{quinn2010feasibility}. The former study uses a mix of cement mortar, acrylics, and rubber lattice as the covering material, while the latter employs Gore-Tex fabric to construct a waterproof cavity for the node. In addition, experimental results from \cite{chang2012implementing} indicate that with suitable encapsulation, the transmission range of the ZigBee wireless module can reach 12\,m and 19\,m when covered by a 30\,cm concrete block with and without a steel beam, respectively.

In the study conducted by Gong et al. \cite{gong2022empowering}, the possibility of powering embedded sensor nodes using acoustic waves instead of RF energy is investigated. In their research, a piezoelectric receiver was employed either to harvest energy or reflect the incident acoustic wave for backscattering communication. Given that the attenuation of mechanical waves propagating through concrete is lower than that of radio waves, piezoelectric receivers can be integrated into thicker concrete structures compared to RF-SNs. However, the acoustic transmitter in \cite{gong2022empowering} must be in direct contact with the concrete surface. If not, a significant path loss would occur at the air-concrete interface. Moreover, to generate acoustic waves with enough strength to power the sensor node, the piezoelectric transmitter would consume substantial power. Consequently, these requirements make it inconvenient for mobile SHM.

\subsection{Motivations}
\label{subsec:Mov}
As introduced in \cite{shen2022experimental} and \cite{qin2009coupled}, periodically measuring the internal temperature and humidity of concrete is vital for assessing material and structural degradation. However, the literature review shows that most existing research on in-situ SHM concentrates on the study of propagation characteristics of RF signals in reinforced concrete. In these studies, the communication between embedded sensor nodes and external receivers is dependent on an active wireless module. While this can provide a reliable link for data transfer, it also results in high power consumption at the node end, thereby posing difficulty for long-term monitoring.

The aforementioned challenges inspire us to develop a more effective solution for sustained SHM. This paper will explore the possibility of powering embedded sensor nodes via a mobile RF transmitter. Depending on the depth of the sensor nodes embedded within the concrete, the RF-SN can transmit data at a high rate using an active wireless module or lower energy consumption using backscattering communication. This new approach allows for sensor deployment in thick concrete structures, delivering consistent evaluations of structural health in a cost-efficient manner.

\section{Application Scenarios}
\label{subsec:App}

RF-SNs are typically designed to assess the condition of concrete structures, including load-bearing walls, dams, and overpasses. In the context of residential construction, a typical thickness of an inner concrete layer is approximately 15\,cm for load-bearing walls~\cite{bida2018advances, lahdensivu2012durability}. As such, the sensor node embedded within the wall will not exceed a depth of 7.5\,cm. Under these circumstances, the active RF-SN is preferred as it is capable of gathering enough radio energy from a mobile radio transmitter for high-speed data transmission.

\begin{figure}[htb]
\centerline{\includegraphics[width=8.7cm]{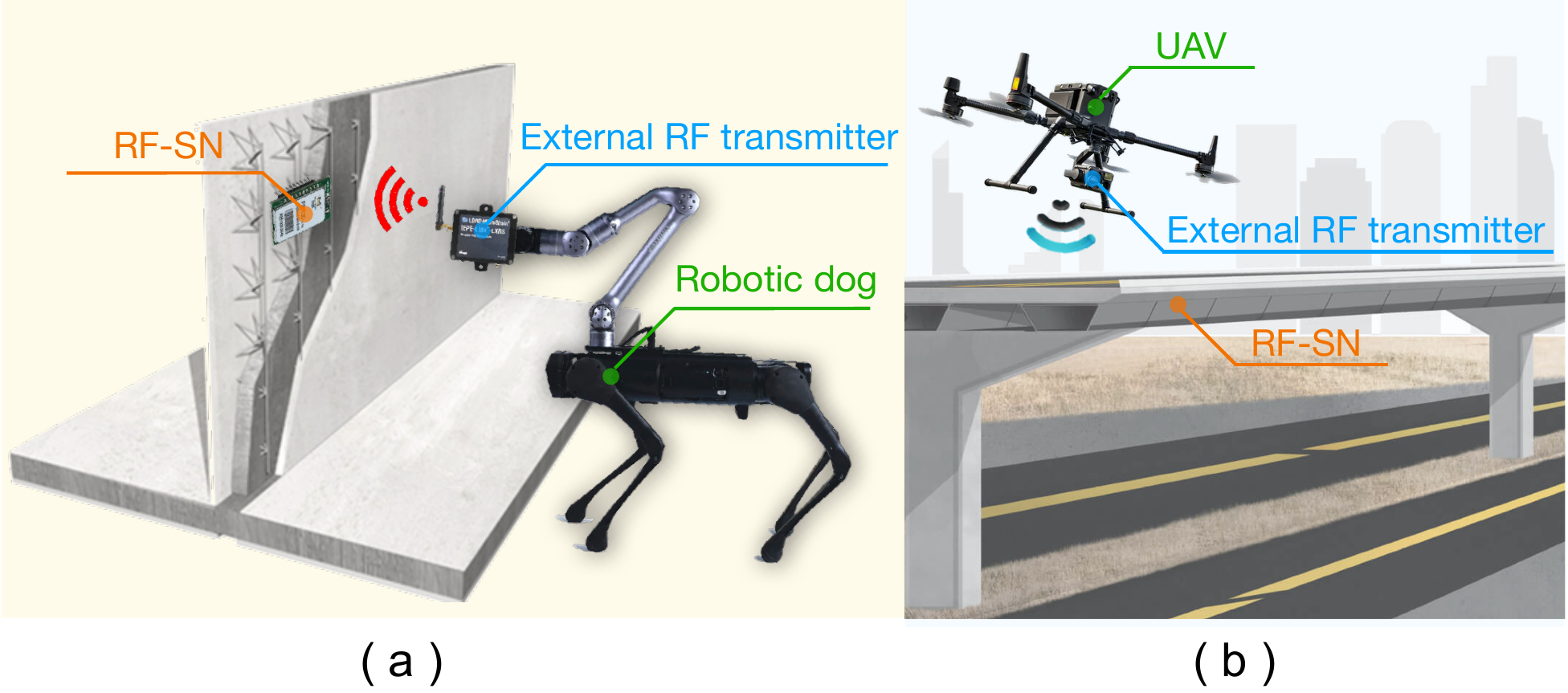}}
  \caption{Application scenarios of RF-SNs with mobile platform. (a) Health check of bearing wall with active RF-SNs. (b) Overpass health monitoring with passive RF-SNs.}\label{fig:Scenario}
\end{figure}

In order to enhance both the efficiency and scalability of active RF-SNs for indoor applications, a robotic dog could be utilized to carry the external RF transmitter and data logger, as illustrated in Fig.~\!\ref{fig:Scenario}(a). The precise positioning of the transmitter can be fine-tuned using the mechanical arm. Given that the placement of each RF-SN is known beforehand, the robotic dog can methodically traverse the designated area to gather sensor data regularly, thereby enabling long-term monitoring.

The dams and overpasses are much thicker than load-bearing walls. Therefore, it is necessary to place the RF-SN much deeper into the concrete to assess the health of these structures. In this case, the sensor node might not receive sufficient energy from the external RF transmitter for active wireless data transmission. Hence, the choice will lean towards a passive RF-SN, which only consumes microwatts of power, thousands of times lower than the active RF-SN for sensing and communication.

When monitoring the health of large outdoor concrete structures, the passive RF-SNs might be placed away from the ground. In order to efficiently charge the node and collect data, a drone can serve as the mobile platform to carry the external RF transmitter and data receiver, as illustrated in Fig.~\!\ref{fig:Scenario}(b). 

In our experiment, we use a Unitree Go 1 robotic dog and a K1 mechanical arm as the primary carrier of the system for indoor testing. However, the total weight of the receiving system, which includes a portable battery, an RF transmitter, and a data receiver, is light --- under 0.5\,kg. Therefore, it can be easily mounted on an industrial drone such as the DJI Matrice 600 Pro, which has a payload capacity of 6\,kg, for outdoor applications~\cite{dji2018matrice}.

\section{Radio waves in Concrete}
\label{sec:RfEng}
In this section, the impact of concrete on the reflection property of the RF antenna will be scrutinized through experiments. Following this, we evaluate the path loss of radio energy transitioning from air to concrete with varying levels of moisture content.

\subsection{Changes of Reflection Property}
\label{subsec:ImpCoff}

In practice, sensor nodes are typically placed within concrete cavities for waterproofing~\cite{chang2012implementing}. By doing this, direct contact between the antenna and the concrete material is avoided, thereby ensuring that the electromagnetic properties of the antenna align with those experienced in an air-filled environment. However, the cavities may potentially undermine the structural integrity of the reinforced concrete. This concern is magnified when there is a need for deploying more than one node to achieve multi-point monitoring.

To tackle the above issue, a feasible strategy involves printing the antenna on the underside of the circuit board, which is subsequently sealed with epoxy resin. Following this, the sensor nodes can be directly incorporated into the concrete. This method eliminates the necessity for constructing cavities, hence facilitating cost-effective sensor deployment. Nonetheless, this approach may present certain compromises, particularly in terms of potential reductions in energy harvesting efficiency and data rate as the concrete may change the properties of the antenna.

Generally, the permittivity of concrete is significantly higher than that of air. Consequently, at a given frequency, the wavelength of a radio signal propagating through concrete will be shorter compared to that in air. Under these circumstances, if an antenna, originally designed to operate in an air environment, is directly embedded in concrete, it may lead to an impedance mismatch, which will in turn affect the performance of energy reception and data transmission. 

\begin{figure}[htb]
\centerline{\includegraphics[width=5.8cm]{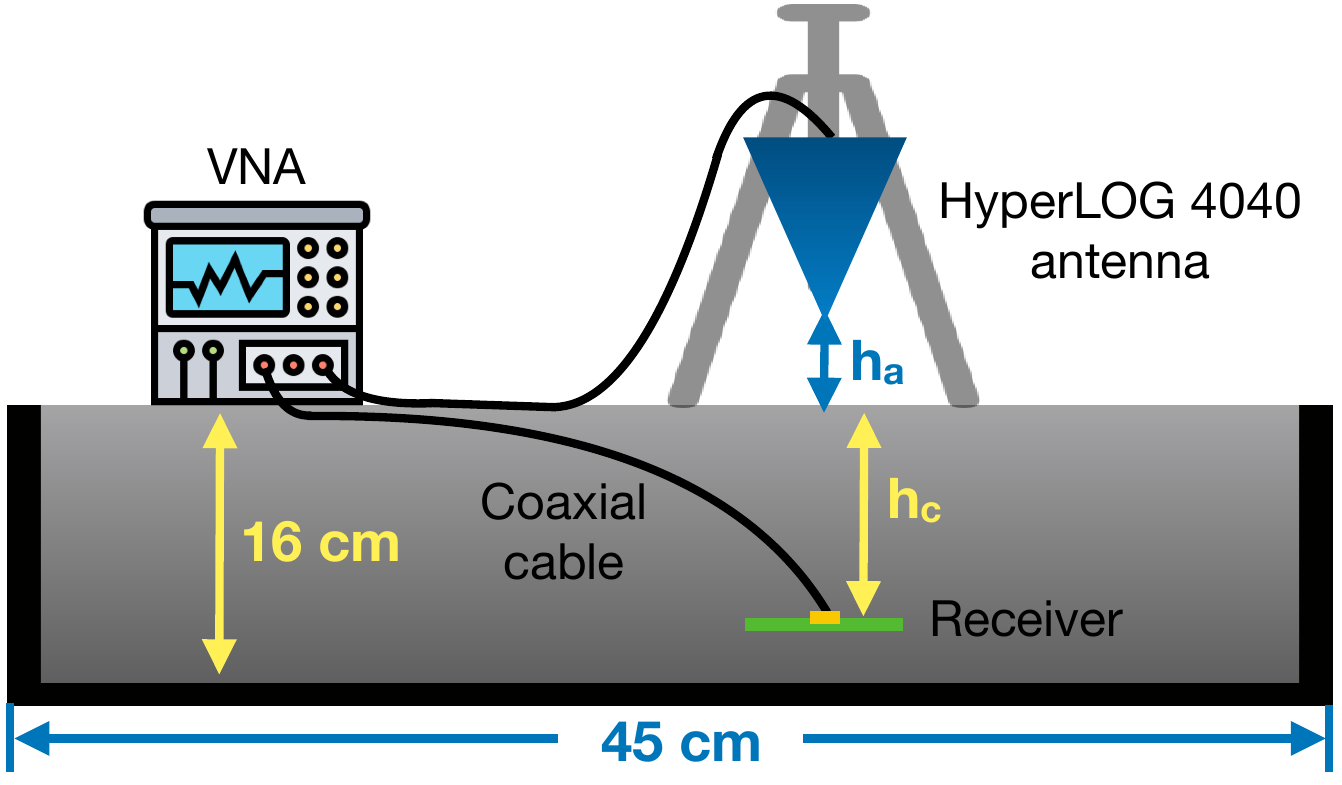}}
  \caption{$S_{11}$ and $S_{21}$ measurements.}\label{fig:ExpRfEng1}
\end{figure}

To give insight into the above conclusion, we carried out a series of experiments to measure the scattering parameters (S-parameters) between an RF transmitter placed in the air and a receiver embedded within concrete. The details of the experimental setup are shown in Fig.~\!\ref{fig:ExpRfEng1} and the mix code of the concrete is listed in Table~\!\ref{tab:ingCon}, where the 28-day characteristic compressive strength of the concrete is 3500\,psi.

As shown in Fig.~\!\ref{fig:ExpRfEng1}, in the experiments, two dipole antennas were embedded at different depths (represented by $h_c$) within a concrete block to receive RF energy. The dimensions of each concrete block were 5\,cm$\times$16\,cm$\times$45\,cm. The receiving antennas were narrowband, with operating frequencies in the air of 915\,MHz and 2.4\,GHz, respectively. The S-parameters were measured via a Keysight N9923A RF vector network analyzer (VNA). To ensure that radio waves could only enter the concrete block from the top, each block was encased in a metal mold. All antennas and equipments are interconnected using coaxial cable to mitigate any potential interferences.

\begin{table}[htp]
\scriptsize
\centering
\caption{Mix code of concrete in experiments}
\label{tab:ingCon}
\begin{tabular}{|
>{\columncolor[HTML]{C0C0C0}}c |
>{\columncolor[HTML]{EFEFEF}}c |
>{\columncolor[HTML]{EFEFEF}}c |
>{\columncolor[HTML]{EFEFEF}}c |
>{\columncolor[HTML]{EFEFEF}}c |
>{\columncolor[HTML]{EFEFEF}}c |
>{\columncolor[HTML]{EFEFEF}}c |}
\hline
Material  & Cement & Fly ash & Crushed stone & Sand & Air & Water \\ \hline
Vol. (\%) & 8.0             & 3.2     & 41.7               & 27.8 & 4.5 & 14.9  \\ \hline
\end{tabular}
\end{table}

\begin{figure}[htb]
\centerline{\includegraphics[width=8.3cm]{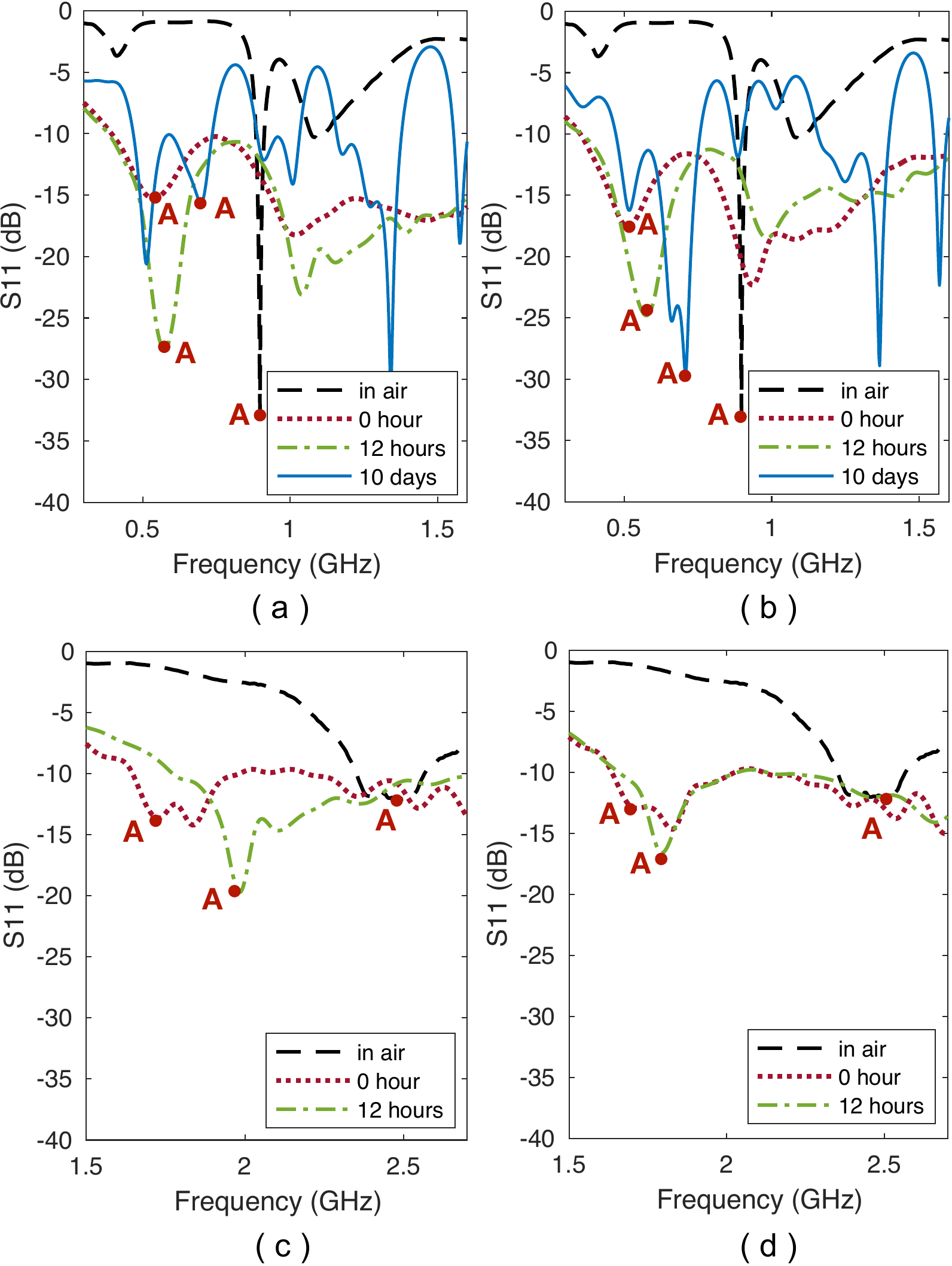}}
  \caption{$S_{11}$ measurements for dipole antennas at various depths and moistures in concrete. (a) $h_c\!=\!3.5$\,cm and $f_r\!=\!915$\,MHz. (b) $h_c\!=\!10$\,cm and $f_r\!=\!915$\,MHz. (c) $h_c\!=\!3.5$\,cm and $f_r\!=\!2.4$\,GHz. (d) $h_c\!=\!10$\,cm and $f_r\!=\!2.4$\,GHz.}\label{fig:S11Mea}
\end{figure}

Fig.~\!\ref{fig:S11Mea} presents the reflection property of the 915\,MHz antenna through $S_{11}$ parameter, measured at varying depths and moisture levels within concrete blocks, where $f_r$ is the resonant frequency of the antenna in the air. The legend of each curve signifies the time interval between the placement of the concrete and the experiment. As depicted in the figure, two significant changes occur when the antenna is directly embedded within the concrete: (a) The $S_{11}$ parameter decreases across the entire frequency band. (b) The antenna's resonant frequency shifts toward a lower frequency as the moisture content of the concrete increases.

As illustrated in Fig.~\!\ref{fig:S11Mea}\,(a), when the antenna is exposed to the air, the black dotted curve displays a deep dip, where the $S_{11}$ value is as low as $-$32.6\,dB at 898\,MHz. However, if the antenna is embedded in the concrete block at a depth of 3.5\,cm ($h_c\!=\!3.5$\,cm), the $S_{11}$ parameter across different frequencies exhibits a "flatter" profile, a pattern that becomes especially noticeable when the concrete becomes wet. Specifically, right after the concrete is poured and its moisture content peaks, the $S_{11}$ measurements (red dotted line in Fig.~\!\ref{fig:S11Mea}\,(a)) taken from 300\,MHz to 1.6\,GHz fall between $-$8\,dB and $-$18\,dB. The flattened reflection property suggests that within the concrete, a narrowband antenna actually displays a broader bandwidth than it does in the air. The underlying reason for this is the inherent nature of concrete as a lossy medium, which absorbs electromagnetic energy, thereby leading to less pronounced antenna resonances. Therefore, although the antenna exhibits a wider bandwidth in the concrete, the overall efficiency of the antenna is likely to decrease due to the absorption of the radio energy in a dielectric material. 

Upon mixing the concrete, the hydration process begins, where water and cement particles combine to create chemical bonds. This reaction precipitates a rapid reduction in moisture content within the first few hours after setting the concrete block. As the processes of hydration and evaporation progress, the electromagnetic energy passing through the concrete experiences less loss, resulting in more distinguishable reflections of the antenna across different frequencies. For example, 12 hours after the concrete has been placed, the $S_{11}$ parameter (green dotted line in Fig.~\!\ref{fig:S11Mea}\,(a)) exhibits a notable fluctuation, ranging between $-$8 dB and $-$27 dB. This variation is substantially more pronounced than that seen in freshly poured concrete. Moreover, a comparison of Fig.~\!\ref{fig:S11Mea}\,(a) and Fig.~\!\ref{fig:S11Mea}\,(b) reveals that $S_{11}$ values measured at the same moisture content but at different depths are remarkably similar, indicating that the antenna's depth within the concrete barely affects the reflective properties.

By comparing Fig.~\!\ref{fig:S11Mea}\,(a) and Fig.~\!\ref{fig:S11Mea}\,(c), it can be observed that the $S_{11}$ values are further flattened by the concrete when using the antenna with a higher resonant frequency. For instance, after embedding the 2.4\,GHz antenna at a depth of 3.5\,cm in freshly poured concrete, the $S_{11}$ values measured from 1.6\,GHz to 2.7\,GHz (red dotted line in Fig.~\!\ref{fig:S11Mea}\,(c)) fall between $-$8.7\,dB and $-$14.1\,dB. Such a range is much flatter than that measured with the 915\,MHz antenna. This is because wet concrete absorbs more energy at higher frequencies than at lower ones. Similar patterns can be observed by comparing Fig.~\!\ref{fig:S11Mea}\,(b) with Fig.~\!\ref{fig:S11Mea}\,(d).

Besides flattening the reflection coefficient, the concrete material also alters the antenna's resonant frequency. As marked by point A in Fig.~\!\ref{fig:S11Mea}\,(b), when exposed to the air, the resonant frequency of the antenna sits at 898\,MHz. However, this frequency shifts to 510\,MHz when the antenna is embedded within a freshly poured concrete block. The shift occurs because concrete typically has a higher permittivity than air, leading to dielectric loading when the antenna is surrounded by concrete.

According to electromagnetic field theory, the relationship between the speed of an electromagnetic wave and the permittivity of the medium is expressed as follows:
\begin{equation}
\label{eq:kf04}
  v_c =\displaystyle\frac{C_0}{\sqrt{\varepsilon_r \,\mu_r}},
\end{equation}
where $v_c$ and $C_0$ are the speeds of electromagnetic in concrete and vacuum, respectively; $\varepsilon_r\!>\!1$ is the relative permittivity of the concrete, and $\mu_r\approx 1$ is the relative permeability of the concrete.

As per equation (\ref{eq:kf04}), given a frequency, the wavelength of an electromagnetic wave in concrete will be significantly shorter than in the air. This explains the shift of point A to the left in Fig.~\!\ref{fig:S11Mea}\,(b) when the concrete is wet. Eventually, as the moisture content reduces, $\varepsilon_r$ of the concrete decreases and the resonant frequency of the antenna gradually increases. For instance, 12 hours and 10 days after the concrete has been poured, the antenna's resonant frequencies rise to 575\,MHz and 706\,MHz, respectively. Similar results can be observed in Fig.~\!\ref{fig:S11Mea}\,(a), where the antenna is buried at a shallower depth within the concrete.

After switching to the 2.4\,GHz antenna, the shift in resonant frequency is less pronounced. This is due to the fact the permittivity of concrete is inversely proportional to the frequency of the electromagnetic waves~\cite{robert1998dielectric}. As a consequence, the proportion of wavelength shortening at high-frequency electromagnetic waves in concrete is lesser than that at low frequency. Particularly, for the 915\,MHz antenna placed at a depth of 10\,cm in the freshly poured concrete block (red dotted line in Fig.~\!\ref{fig:S11Mea}\,(b)), its resonant frequency moves from 898\,MHz to 503\,MHz, the frequency shift of which is over 44\%. In this case, the permittivity value of the concrete calculated via (\ref{eq:kf04}) is $\varepsilon_r \!=\!3.2$. In contrast, for the 2.4\,GHz antenna measured under the same conditions (red dotted line in Fig.~\!\ref{fig:S11Mea}\,(d)), its resonant frequency moves from 2.4\,GHz to 1.7\,GHz. The frequency shift in this scenario is only 29\% as the permittivity of the concrete at 2.4\,GHz decreases to 2.

Based on the above observations, it can be inferred that for maximizing energy harvesting efficiency, the frequency of the radio energy should not be static. Instead, it can be dynamically adjusted in accordance with the moisture content in the concrete. Generally, the antenna's resonant frequency decreases after being embedded in the concrete. Hence, the RF transmitter should also emit a lower frequency than in the open-air environment to align with the receiving antenna, especially in a low-frequency scenario.

\subsection{Path Loss of RF Energy in Concrete}
\label{subsec:Pathloss}
To evaluate the path loss between the RF transmitter and receiver, a log-periodic antenna, HyperLOG 4040, is employed to transmit the signal. This antenna provides a gain of 4\,dBi in the range of 400\,MHz to 4\,GHz~\cite{aaronia2014logper}. The distance between the top of the HyperLOG 4040 and the surface of the concrete block, denoted as $h_a$, is maintained at 36\,cm throughout the experiment, as illustrated in Fig.~\!\ref{fig:ExpRfEng1}. The transmission antenna is positioned perpendicular to the surface of the concrete. Hence, the angle of incidence approximates to zero.

\begin{figure}[htb]
\centerline{\includegraphics[width=8.3cm]{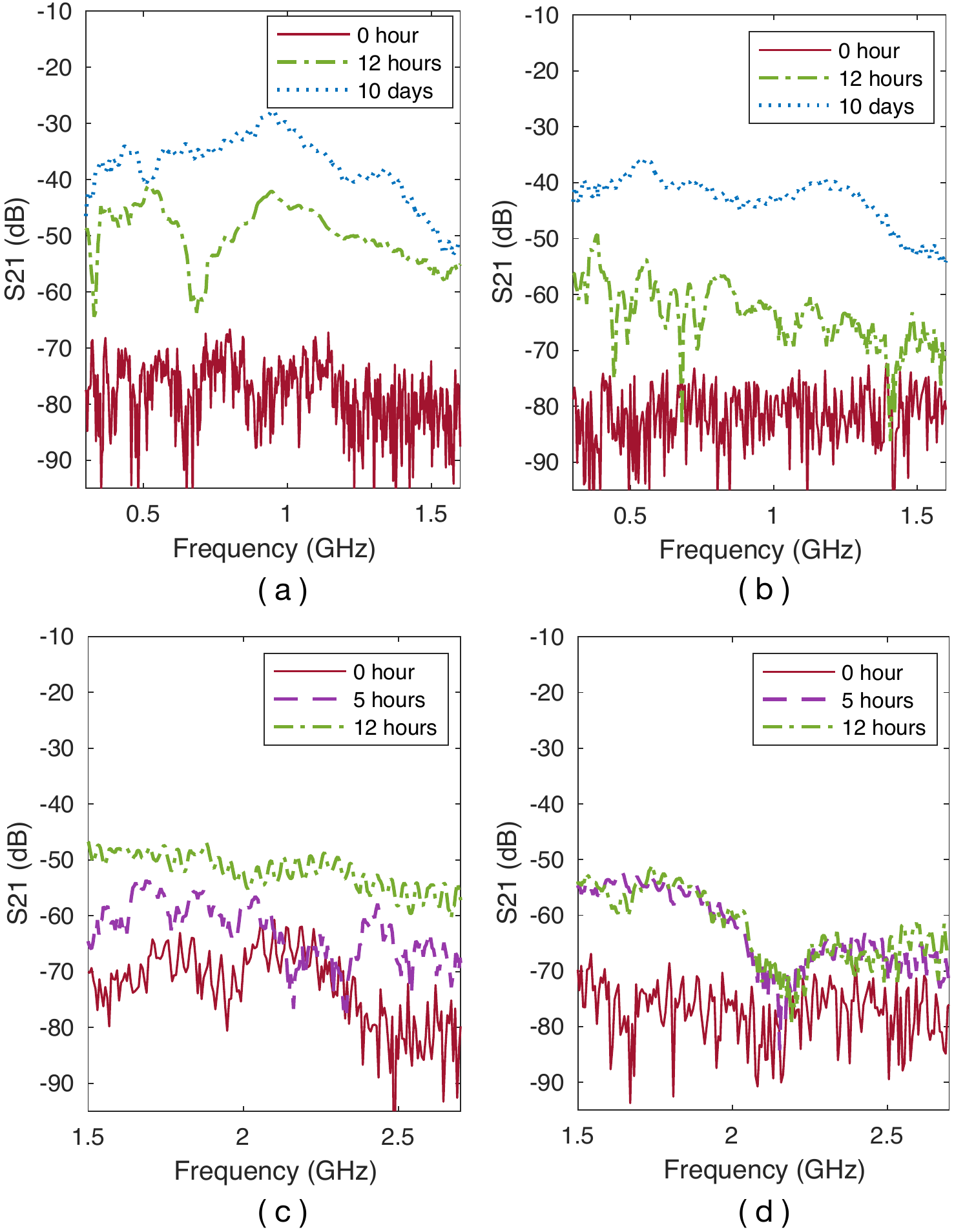}}
  \caption{$S_{21}$ measurements for dipole antennas at various depths and moistures in concrete. (a) $h_c\!=\!3.5$\,cm and $f_r\!=\!915$\,MHz. (b) $h_c\!=\!10$\,cm and $f_r\!=\!915$\,MHz. (c) $h_c\!=\!3.5$\,cm and $f_r\!=\!2.4$\,GHz. (d) $h_c\!=\!10$\,cm and $f_r\!=\!2.4$\,GHz.}\label{fig:S21Mea}
\end{figure}

Fig.~\!\ref{fig:S21Mea} displays the $S_{21}$ parameter measured at different depths and moisture contents within concrete blocks. As demonstrated in the figure, the moisture content significantly influences the path loss of RF energy. For instance, with the 915\,MHz antenna embedded at a depth of 3.5\,cm (i.e., $f_r\!=\!915$\,MHz and $h_c\!=\!3.5$\,cm), the path loss of RF energy is as high as 82\,dB around 915\,MHz in freshly poured concrete, as depicted in Fig.~\!\ref{fig:S21Mea}\,(a). However, this value decreases to 43\,dB and 29\,dB after the concrete has been set for 12 hours and 10 days, respectively.

By comparing Fig.~\!\ref{fig:S21Mea}\,(a) with Fig.~\!\ref{fig:S21Mea}\,(b), we can eliminate the reflection loss of RF energy at the air-concrete interface, as well as the effects of antenna characteristics, thereby allowing us to estimate the propagation attenuation of the radio wave in concrete. Specifically, when the antenna is buried 10\,cm deep in the concrete, the $S_{21}$ measured in the dry concrete is $-$43.6\,dB at 915\,MHz, 14.6\,dB lower than the value measured at $h_c\!=\!3.5$\,cm. Accordingly, the propagation loss of 915\,MHz radio energy in a dry concrete block amounts to approximately 2.2\,dB/cm. This value increases to 3\,dB/cm in wet concrete 12 hours after it is poured.

Upon comparing Fig.~\!\ref{fig:S21Mea}\,(b) and Fig.~\!\ref{fig:S21Mea}\,(d), we observe an increase in the path loss of RF energy as the frequency rises. More specifically, as shown in Fig.~\!\ref{fig:S21Mea}\,(d), when the burial depth of the antenna is 10\,cm and the concrete has been setting for 12 hours, $S_{21}$ obtained at 2.4\,GHz can achieve $-$69\,dB, which is 7\,dB lower than the measurement at 915\,MHz. Moreover, if the burial depth is reduced to 3.5\,cm, this difference further broadens to 13\,dB.

From Fig.~\!\ref{fig:S21Mea}, we observe that $S_{21}$ does not show an appreciable peak at the resonant frequency of the receiving antenna, especially in wet concrete. In simpler terms, the RF energy received at the antenna's resonant frequency is not substantially greater than that received at other surrounding frequencies. This outcome is a result of the inherent properties of concrete as a lossy medium, which smooths the reflection profile of the receiving antenna across frequencies, as elaborated in Section~\!\ref{subsec:ImpCoff}. Based on this observation, we can appropriately raise the frequency of RF waves without hurting the energy harvesting rate. This in turn allows us to reduce the size of both the transmitting and receiving antennas, thereby enabling more flexible sensor deployment.

\section{Architecture of Active RF-SN}
\label{sec:ActArch}

\begin{figure}[htb]
\centerline{\includegraphics[width=8.5cm]{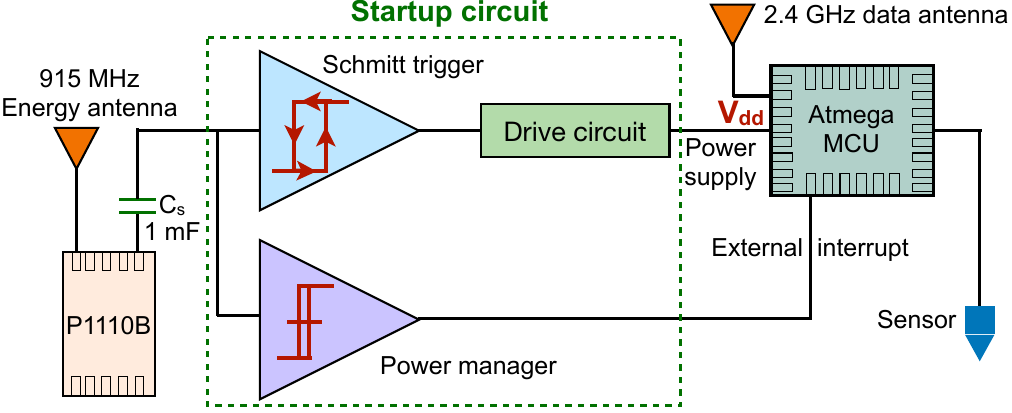}}
  \caption{Architecture of the active RF-SN, where $C_s$ is the energy storage capacitor and $V_{dd}$ is the supply voltage for the MCU.}\label{fig:actCir}
\end{figure}

Fig.~\!\ref{fig:actCir} presents the architecture of the active RF-SN that we developed for health monitoring in thin concrete structures. Here, we select the 915\,MHz frequency band for radio energy delivery based on two primary reasons: (a) It provides an ideal equilibrium between the size of the antenna and the path loss of radio energy from the air to concrete. (b) A wide array of commercial components including power amplifiers, envelope detectors, and bandpass filters are designed for this frequency range, greatly reducing the cost of node development. 

As shown in Fig.~\!\ref{fig:actCir}, a Powercast P1110B chipset is utilized to convert RF energy from alternating current (AC) to direct current (DC)~\cite{powercast2018p1110b}. The harvested energy is then stored in a 1\,mF capacitor, supplying the node with the requisite power. In an effort to minimize power consumption, the Microchip ATmega256RFR2 microcontroller unit (MCU) is chosen for data sampling and communication. This MCU presents two compelling features. Firstly, it comes with an integrated 2.4\,GHz module, enabling it to conveniently transmit data via the 802.15.4 Zigbee protocol. Secondly, ATmega256RFR2 has a remarkably low current consumption, which is only 650 nA in deep sleep mode~\cite{microchip2016atmel}, rendering it an attractive option for low-power applications.

It is crucial to emphasize that the MCU should not be directly linked to the energy storage capacitor; otherwise, the controller would consistently draw substantial energy even before the capacitor attains the minimum voltage required for the proper function of the MCU. Furthermore, the power consumption of the MCU during its startup phase far surpasses the energy harvesting rate. This imbalance could lead to startup failure if the MCU were to drain energy from the capacitor directly.

To address the aforementioned problem, we have developed a startup circuit specifically for the RF-SN, as depicted in Fig.~\!\ref{fig:actCir}. This circuit is composed of three primary components: a Schmitt trigger, a current driver, and a power manager. The Schmitt trigger is proposed to postpone the startup voltage of the ATmega256RFR2, elevating it from 1.8\,V to 3.2\,V. This strategy enables the capacitor to gather sufficient energy for a reliable start-up of the MCU. The current driver is essentially a modified RC circuit, serving as a current extender. It provides the necessary current burst needed to power the MCU, while preserving the energy after finishing the startup process. More details on the current driver can refer to our previous research \cite{luo2023uav}.

After the initial startup and first-round data transmission, there is a notable decrease in the charge of the energy storage capacitor. Consequently, it becomes essential for the MCU to swiftly transition into sleep mode, allowing the capacitor short respite to recharge. Concurrently, the comparator in the power manager monitors the voltage level of the energy storage capacitor. When the voltage reaches a pre-determined upper threshold (specifically 2.6\,V in our configuration), the comparator triggers an external interrupt. This action rouses the MCU, enabling it to sense and send data. At the same time, the MCU assesses its supply voltage and reverts to sleep mode once the voltage descends to a lower threshold (set at 2.3\,V in our implementation).

\begin{figure}[htb]
\centerline{\includegraphics[width=8.5cm]{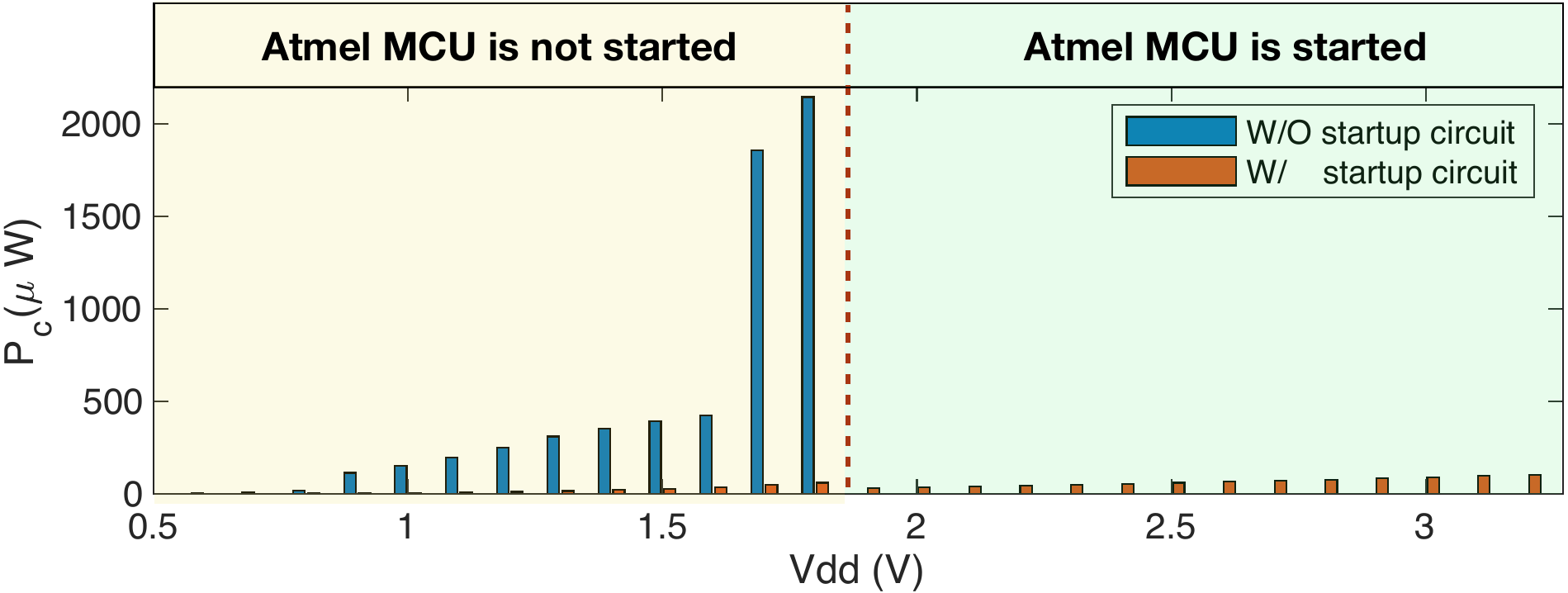}}
  \caption{Power consumption of the active RF-SN with and without startup circuit.}\label{fig:WOStartup}
\end{figure}

Fig.~\!\ref{fig:WOStartup} compares the power consumption of the active RF-SN operating both with and without a startup circuit, where $P_c$ represents the power consumption of the RF-SN. As evident from the figure, in the absence of the startup circuit, the power consumption of the RF-SN is strikingly high. This situation is exacerbated when $V_{dd}$ approaches 1.8\,V, the designated startup voltage for the ATmega256RFR2. More specifically, at a comparatively low supply voltage of 0.6\,V, the power consumption of the MCU that is not yet started is remarkably low, resting at approximately 3.1\,$\mu$W. However, as the supply voltage rises to 1.8\,V, the power consumption experiences a dramatic increase, reaching 2.1\,mW. As a result, the incident power of RF energy must attain a minimum of 3.5\,mW (equivalent to 5.4\,dBm). This requirement ensures a balance between energy income and expenditure for P1110B, which provides a 60\% energy conversion efficiency when operating at an incident power of 5\,dBm~\cite{powercast2018p1110b}.

Based on the $S_{21}$ parameter measured in Section~\!\ref{subsec:Pathloss}, the pass loss for 915\,MHz energy is found to be $-$43.6\,dB. As we are about to demonstrate in Table~\!\ref{tab:strCon} of Section~\!\ref{subsec:related}, bringing the transmitting antenna closer to the concrete can cut this value by at least 11\,dB. Even under these circumstances, an RF transmitter with an EIRP of 36\,dBm (equivalent to 4\,W), a power that complies with FCC limitations, can only deliver energy with an intensity of 3.2\,dBm to the RF-SN embedded 10\,cm deep in a dry concrete structure. This value is lower than the minimum requirement, which is 5.4\,dBm calculated above, to power the MCU without the startup circuit.

As illustrated in Fig.~\!\ref{fig:WOStartup}, the power consumption of the RF-SN drastically decreases to 61\,$\mu$W with the assistance of the startup circuit. This is only a 2.9\% of the power consumption when operating without the startup circuit. Specifically, when the supply voltage is below 1.8\,V, the Schmitt trigger's output is zero, resulting in the MCU being completely shut down. In such a scenario, only the Schmitt circuit and the power manager consume a small amount of energy. This substantially conserves energy consumption and the minimum incident power necessary for the active RF-SN to operate properly.

\section{Architecture of Passive RF-SN}
\label{sec:PassArch}
This section delves into the architecture of the passive RF-SN. We first introduce the circuit design and follow it up with a discussion on enhancing the communication performance of the passive RF-SN from the perspective of signal modulation.

\subsection{Circuit Structure}
\label{subsec:CirStr}
The passive RF-SN is designed to operate within thick concrete structures. Therefore, the sensitivity of the RF energy harvester \footnote{The sensitivity of energy harvester is characterized by the minimum intensity of incident power required to charge the energy storage capacitor to 1.8\,V, which is necessary to successfully  activate the MCU.} needs to exceed the active one. To achieve this, we select a Powercast PCC110 RF-DC converter and a PCC210 voltage booster to construct the energy harvester~\cite{powercast2018pcc110}. This combination enables the RF-SN to collect sufficient energy even when the intensity of the incident power falls below $-$8\,dBm. 

Beyond enhancing the energy harvester's sensitivity, the passive RF-SN is expected to be more power-efficient than the active one. To achieve this, the sensor nodes are intentionally designed not to generate carrier signals for wireless communication --- a procedure that typically involves a high-frequency oscillator and is associated with considerable power consumption. Instead, the RF-SN operates similarly to an RFID tag, modulating the amplitude of the backscattered radio energy sourced from the mobile transmitter. This approach promotes passive data transmission, optimizing energy usage.

\begin{figure}[htb]
\centerline{\includegraphics[width=8.5cm]{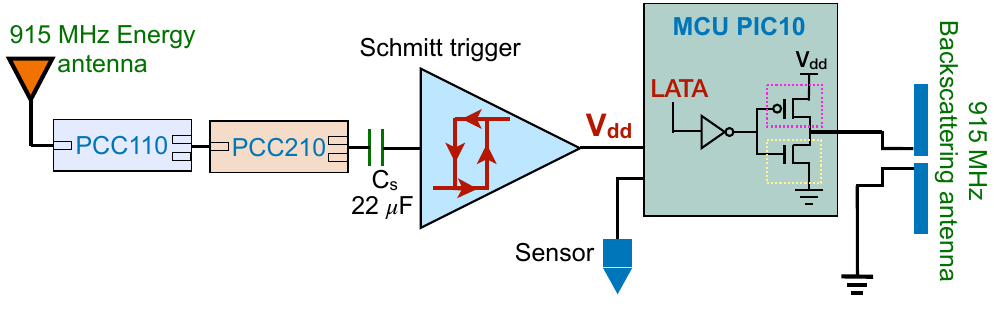}}
  \caption{Architecture of the passive RF-SN.}\label{fig:PassCir}
\end{figure}

As illustrated in Fig.~\!\ref{fig:PassCir}, in an effort to decrease overall energy consumption within the system, the MCU has been substituted with a Microchip PIC10LF322. This replacement only requires 36\,nW and 8.1\,$\mu$W of power to stay in sleep and operational modes\footnote{The power consumption is measured at 32.768\,kHz clock frequency and 1.8\,V supply voltage.}, respectively. To modulate the amplitude of the backscattered signal, a 915\,MHz dipole antenna having the same resonant frequency as the RF energy is connected to a general-purpose input/output (GPIO) port of the MCU.

If writing `1' to the LATA register of the PIC10,  the PMOS inside the GPIO port is switched off and the NMOS is switched on, shorting the antenna. Conversely, writing `0' to the register activates the PMOS and deactivates the NMOS, opening the antenna. In the first scenario, the strength of the scattered wave is the same as the incident wave, while in the second one, it approaches zero. Hence, by assigning different values to the LATA register, the RF-SN can effectively modulate the amplitude of the scattered wave.

\subsection{Signal Structure}
\label{subsec:SigStr}
Given the significant reflection loss of radio energy at the air-concrete interface and the substantial propagation attenuation of the RF signal within concrete material, the backscatter signal produced by the passive RF-SN is usually weak. Therefore, instead of using on-off keying (OOK), a modulation scheme extensively applied in commercial RFIDs, we use a sequence of square chirps for data transmission to enhance the bit error rate (BER) at low signal-to-noise ratio (SNR) levels. 

\begin{figure}[htb]
\centerline{\includegraphics[width=8.5cm]{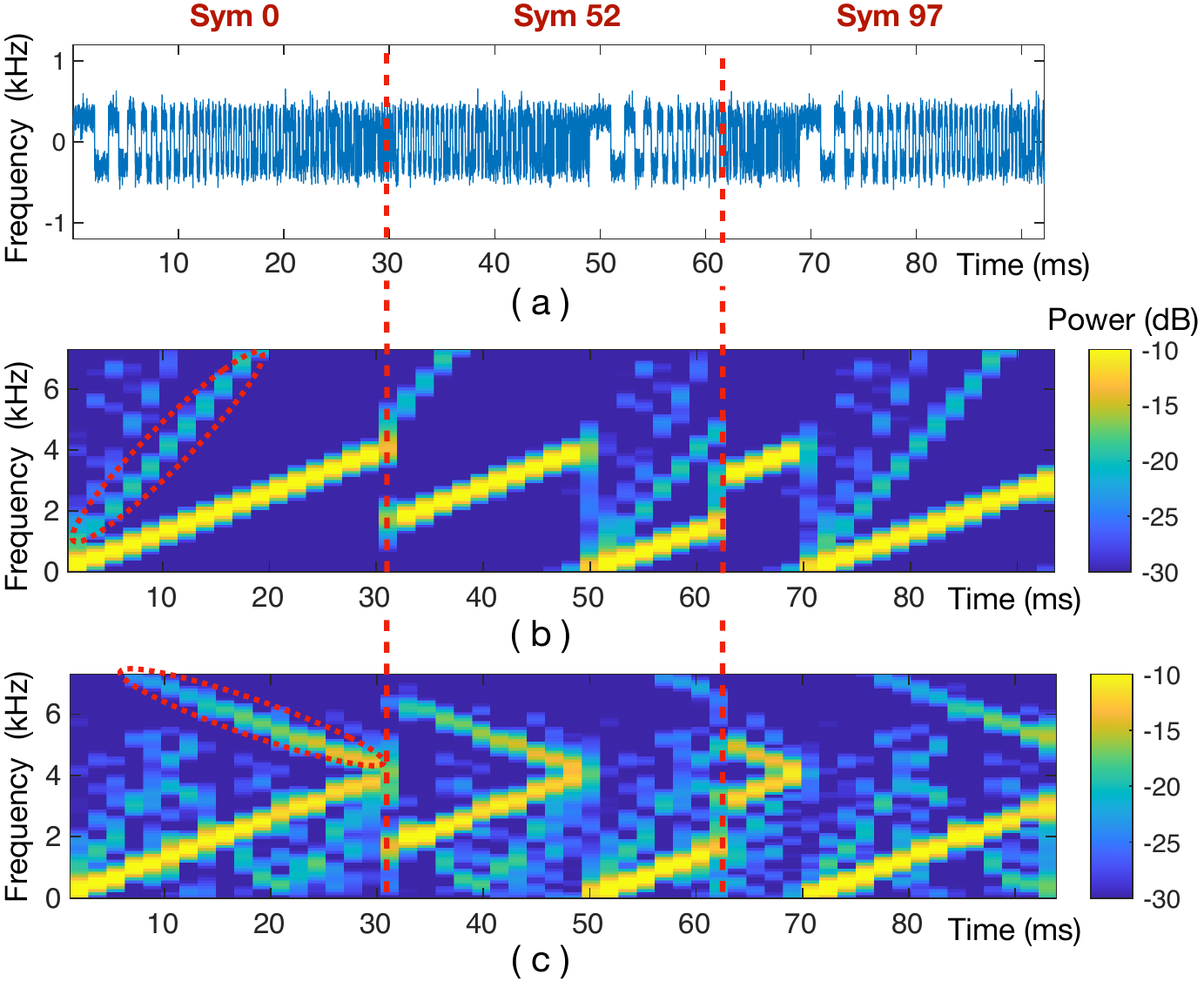}}
  \caption{Characteristics of square chirps,  where SF \!=\! 7. (a) Waveforms of square chirps. (b) Time-frequency spectrum of ideal square chirps. (c) Time-frequency spectrum of real square chirps generated by PIC10 at 32.768\,kHz clock frequency.}\label{fig:CmpSpec}
\end{figure}

Fig.~\!\ref{fig:CmpSpec} provides a sample of three square chirps produced by the passive RF-SN at a clock frequency of 32.768\,kHz and SNR of 9.5\,dB. In the figure, the spreading factor (SF) of the signal is set to 7, representing the number of data bits encoded in a symbol; the duration of each symbol is 31\,ms, while the bandwidth of the signal is 4.1\,kHz.

As depicted in Fig.~\!\ref{fig:CmpSpec}\,(a), even though the bandwidth is consistent across all square chirps, distinct symbols are represented through varying start and end frequencies of the signal. Therefore, the time-frequency diagrams of square chirps display a piecewise linear trajectory over time. This characteristic is similar to linear chirps, which are commonly used in radar and sonar systems. However, unlike a linear chirp whose amplitude varies continuously over time, the GPIO ports of the PIC10 are digital, which can only alternate between the ON and OFF states. This results in a binary amplitude for the square chirps. 

With a binary amplitude, the spectrum of a square chirp at any specific moment encompasses a main frequency and multiple odd harmonic components, as highlighted by the red circle in Fig.~\!\ref{fig:CmpSpec}\,(b) as an example. This occurs because the Fourier transform of a square wave is a sinc function, which has local maxima at odd harmonic frequencies. For data detection, these harmonics are perceived as interference, leading to a reduction in the SNR.

In addition to interference from harmonic frequencies, the quality of the square chirp produced by the passive RF-SN can be further compromised by the low clock rate of the MCU, as it may cause inaccurate toggling timing at the GPIO. Particularly, the PIC10 requires at least one instruction cycle, equivalent to four clock cycles, to toggle the output of the GPIO port. Therefore, the time necessary for the MCU to change the state of the I/O port must be an integer multiple of $4/F_{osc}$, where $F_{osc}$ is the clock's oscillation frequency. 

When the MCU runs at a low clock rate, the square chirp generated by the passive RF-SN might deviate from the ideal square chirp. Fig.~\!\ref{fig:CmpSpec}\,(c) reveals that when the MCU is powered by the internal crystal oscillating at 32.768\,kHz, the spectrum of the actual backscattering signal is significantly noisier than the ideal one, especially at higher frequencies within each chirp. This disparity can adversely affect the performance of signal detection.

It is worth noting that for the passive RF-SN, the maximum bandwidth of the backscattering signal is constrained by the MCU's clock rate. Generally, toggling the output of the I/O port requires 4 clock cycles. This results in a minimum period of 8 clock cycles for the square wave that the PIC10 can generate. In other words, the maximum bandwidth of the square chirp produced by the PIC10 is only 1/8 of the clock frequency. For instance, if the MCU is configured to operate at 1\,MHz, the maximum bandwidth of the chirp signal can reach 125\,kHz. By contrast, the bandwidth will shrink to 4.1\,kHz if the clock frequency is lowered to 32.768\,kHz. However, increasing the oscillation frequency of the clock will consume considerable energy. Therefore, there is a tradeoff between the signal bandwidth and the system power consumption. Employing a low-frequency clock can considerably reduce power consumption, enabling the node to operate in thick concrete. Nevertheless, this reduces the bandwidth, which may cause either a low data rate or a high BER at a low SNR scenario.

\section{Experimental results}
\label{sec:ExpRes}

This section shares the results derived from experiments. We first scrutinize the strength of RF energy that the sensor node can capture within the concrete, and then gauge the charging rate of the RF-SN under various conditions. Following this, we undertake a thorough evaluation of the communication performance of both active and passive RF-SNs.

\subsection{RF Intensity and Charging Rate}
\label{subsec:RFChar}

\subsubsection{RF Intensity}
\begin{figure}[htb]
\centerline{\includegraphics[width=8.5cm]{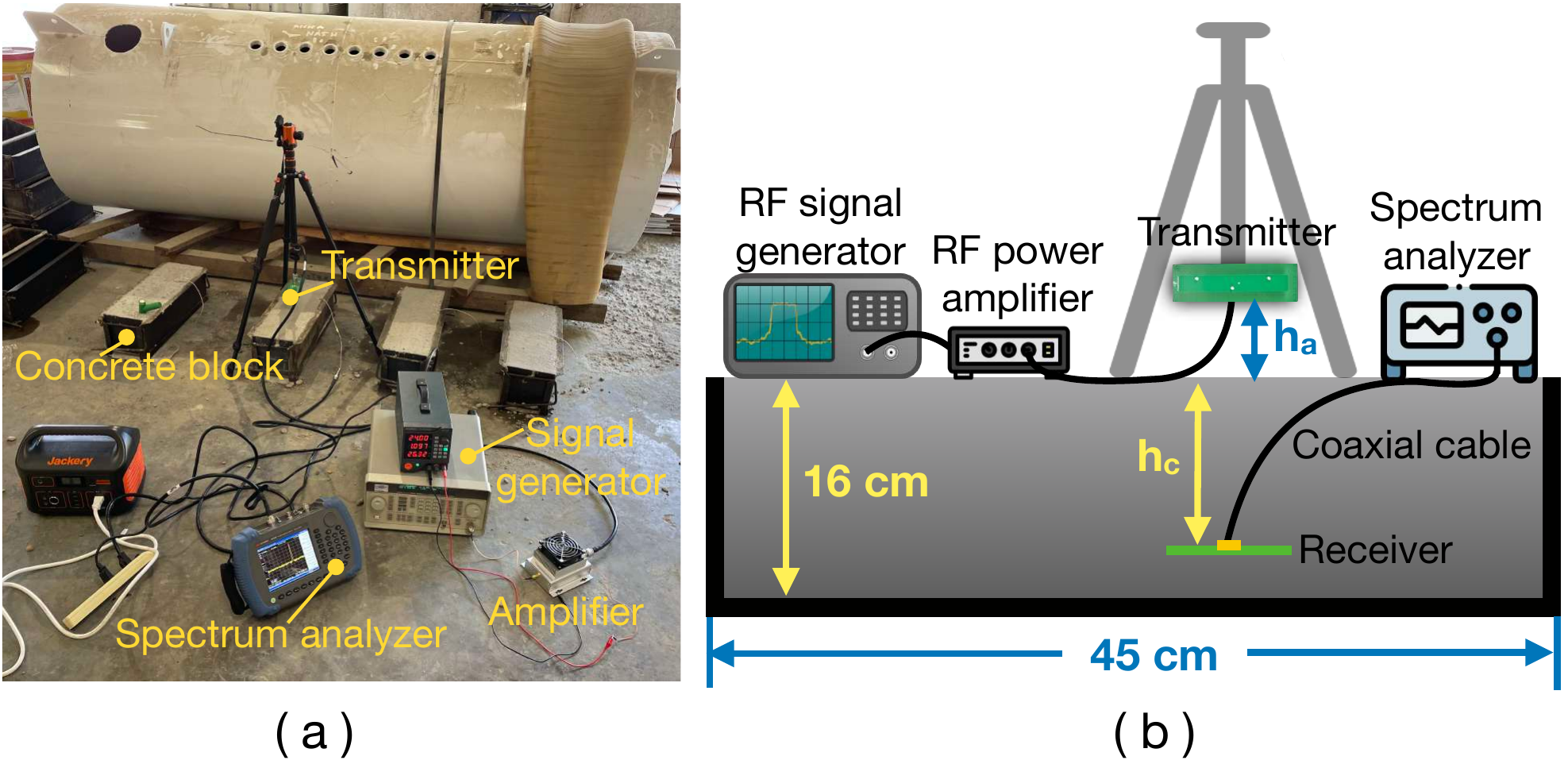}}
  \caption{Intensity of incident RF energy. (a) Experimental setup. (b)  Incident power measurements.}\label{fig:ExpRfEng2}
\end{figure}

Results presented in Section~\!\ref{subsec:Pathloss} indicate a high path loss for 2.4\,GHz radio waves in the concrete. Therefore, we have opted for the 915\,MHz frequency band to carry RF energy thereby striking a balance between antenna size and energy harvesting rate. The experimental setup, as shown in Fig.~\!\ref{fig:ExpRfEng2}, involves a Hewlett-Packard 8648C signal generator that produces a continuous radio signal. This signal is then bolstered by a power amplifier and radiated by a narrowband patch antenna, which provides an 8\,dBi gain in the boresight direction. The distance between the transmission antenna and the surface of the concrete block is $h_a\!=\!$ 6\,cm. This distance is shorter compared to that established in Section~\!\ref{subsec:Pathloss}, where $h_a\!=\!$ 36\,cm. Therefore, the path loss between the transmitter and receiver will be less than the one depicted in Fig.~\!\ref{fig:S21Mea}.

At the receiver end, dipole antennas with a directional gain of 1\,dBi are embedded at varying depths in concrete blocks. Similar to Fig.~\!\ref{fig:ExpRfEng2}, all concrete blocks are enclosed within a metal mold to ensure that the radio waves can only enter from the top. The mix code of the concrete used in experiments can be found in Table~\ref{tab:ingCon}. The strength of the received energy is gauged using an Agilent N9340B spectrum analyzer.

Let $P_r$ represent the intensity of received energy and $d_p$ indicate the depth of the receiving antenna within the concrete. Table~\!\ref{tab:strCon} lists the $P_r$ values measured 7 days after the concrete has been poured. The data highlighted in the colored cell will be cross-referenced with Fig.~\!\ref{fig:Charging} to evaluate the RF-SN's charging rate in different scenarios. In accordance with FCC regulations, the EIRP of transmission energy within the ISM frequency band should not exceed 4\,W, or equivalently, 36\,dBm. Adhering to this restriction, $P_r$ can reach 11.4\,dBm for a receiver situated 3.5\,cm deep within the concrete. This value diminishes to 3.2\,dBm when $d_p$ is expanded to 10\,cm. The active RF-SN we constructed for SHM demonstrates an energy harvesting sensitivity of approximately $-$2.5\,dBm. Thus, a mobile radio transmitter with an EIPR of 36\,dBm is adequate to activate the sensor node embedded 13.5\,cm deep within the concrete.

\begin{table}[htp]
\scriptsize
\centering
\caption{Intensity of incident RF energy in concrete}
\label{tab:strCon}
\begin{tabular}{|
>{\columncolor[HTML]{C0C0C0}}c |
>{\columncolor[HTML]{EFEFEF}}c |
>{\columncolor[HTML]{EFEFEF}}c |
>{\columncolor[HTML]{EFEFEF}}c |c|c|
>{\columncolor[HTML]{EFEFEF}}c |}
\hline
EIRP                            & 9.7   & 13.9  & 17.8                         & \cellcolor[HTML]{EFEFEF}23.6 & \cellcolor[HTML]{EFEFEF}30.7 & 36                        \\ \hline
$P_r$ at $d_p\!=\!$ 3.5\,cm     & -17.0 & -13.4 & \cellcolor[HTML]{FFCE93}-8.1 & \cellcolor[HTML]{CBCEFB}-1.5 & \cellcolor[HTML]{EFEFEF}5.9  & 11.4                        \\ \hline
$P_r$ at $d_p\!=\!$ 6\,cm       & -20.2 & -16.5 & -11.2                        & \cellcolor[HTML]{FFCE93}-4.7 & \cellcolor[HTML]{CBCEFB}2.5  & 8.2                         \\ \hline
$P_r$ at $d_p\!=\!$ 10\,cm      & -24.5 & -20.9 & -16.0                        & \cellcolor[HTML]{EFEFEF}-9.6 & \cellcolor[HTML]{FFCE93}-2.3 & \cellcolor[HTML]{CBCEFB}3.2 \\ \hline
$P_r$ at $d_p\!=\!$ 13.5\,cm & -23   & -19.5 & -14.2                        & \cellcolor[HTML]{FFCE93}-7.3 & \cellcolor[HTML]{CBCEFB}0.25 & 6.0                         \\ \hline
\end{tabular}
\end{table}

An intriguing observation can be made from Table~\!\ref{tab:strCon}: the $P_r$ value measured at a depth of 13.5\,cm surpasses that recorded at 10\,cm. This experiment has been repeated several times, obtaining the same result. Currently, we are unable to provide a clear explanation for this counter-intuitive result. A potential hypothesis could be that the metal mold serves as a reflective cavity, thereby focusing the RF energy at specific locations. However, a more in-depth analysis would necessitate modeling the electromagnetic fields within the metal mold filled with concrete, which falls beyond the scope of this paper.

For the active RF-SN, we configured its MCU clock frequency, wireless transmission power, data rate, and MAC service data unit (MSDU) payload to 8\,MHz, 3.5\,dBm, 1\,Mbps, and 105\,octets, respectively. Under these settings, the node consumes 1.68\,mJ and 177\,$\mu$J energy to wake up the MCU and send each data packet at a supply voltage of 2.3\,V, respectively. According to Table~\!\ref{tab:strCon}, when the EIRP of the transmitter is 32.5\,dBm, the radio energy received by the RF-SN can achieve 2\,dBm. In this case, the active RF-SN can be successfully started and send over 1 kilobyte of data within the first 10 seconds, followed by an additional 1.7 kilobytes every 1.6 seconds of extra charging.

If we opt for the passive RF-SN, the energy harvesting sensitivity is further improved to $-$8.5\,dBm. This enhanced sensitivity enables the sensor node to be situated more deeply within the structure than the active RF-SN, providing the capability to monitor the health of thicker concrete structures, or alternatively, it could reduce the power usage of the mobile RF transmitter to save energy.

\subsubsection{Charging Rate}
While the $S_{21}$ parameter presented in Fig.~\!\ref{fig:S21Mea} and the intensity of incident power listed in Table~\!\ref{tab:strCon} can provide a preliminary estimate of the charging rate for RF-SNs, the results may not be accurate. This discrepancy arises because, during the energy harvesting process, both the startup circuit and the MCU continuously consume energy. This energy consumption is not constant but increases nonlinearly in tandem with the voltage of the energy storage capacitor. As such, experimental measurements for a precise evaluation of the RF-SN's charging rate are essential in real applications.

\begin{figure}[htb]
\centerline{\includegraphics[width=8.5cm]{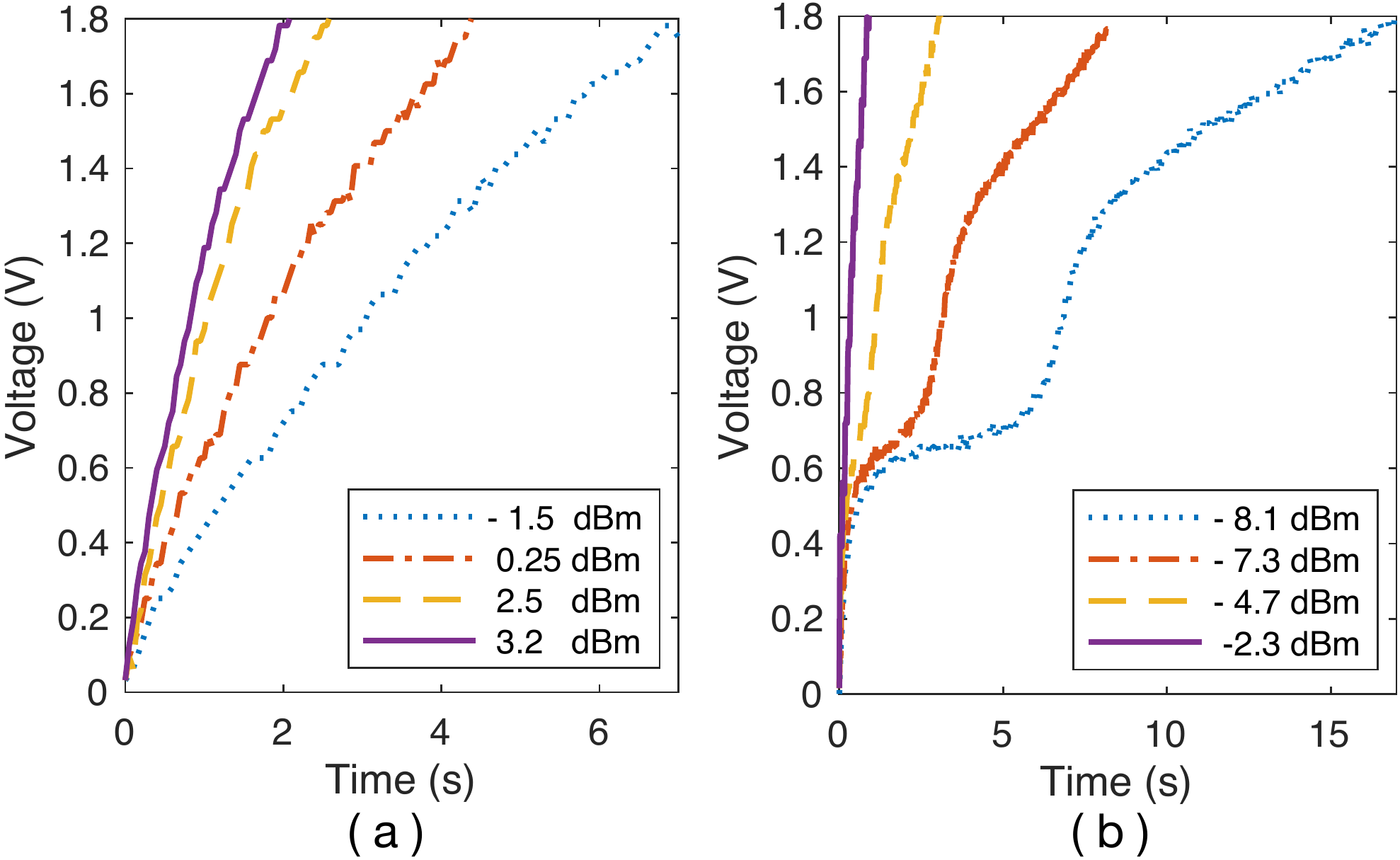}}
  \caption{Charging time of RF-SN at different intensities of incident power. (a) Active RF-SN with 1\,mF energy storage capacitor. (b) Passive RF-SN with 22\,$\mu$F energy storage capacitor.}\label{fig:Charging}
\end{figure}

Fig.~\!\ref{fig:Charging} showcases the RF-SN's charging rates under varying incident power levels. The associated depth and EIRP information are provided by the highlighted cells in Table~\!\ref{tab:strCon}. The active RF-SN, given its high power consumption and the large capacitance of its energy storage capacitor essential for a successful startup, necessitates a considerably higher incident power than its passive counterpart to sufficiently charge the energy storage within a reasonable duration. For instance, an active RF-SN, when embedded 3.5\,cm deep in concrete, is capable of harvesting $-$1.5\,dBm of energy from a transmitter with an EIRP of 23.6\,dBm. Under these circumstances, the energy harvester takes approximately 6.8 seconds to charge the capacitor to 1.8\,V. In contrast, a passive RF-SN situated 13.5\,cm deep in the concrete block only needs $-$7.3\,dBm of incident power from a transmitter with the same EIRP to reach the same capacitor voltage level within a similar charging period.

As depicted in Fig.~\!\ref{fig:Charging}, boosting the incident power can greatly shorten the charging time, particularly when the incident power is already high. Consider the passive RF-SN as an example: when $P_r$ is $-$8.1\,dBm, the charging duration is 17 seconds. Yet, this duration decreases dramatically to 0.9 seconds when $P_r$ is increased to $-$2.3\,dBm. In this situation, despite the incident power only seeing an increase of 3.8 times, the charging time undergoes a dramatic reduction by 94.7\%.

\subsection{Communication Performance of Passive RF-SN}
\label{subsec:ComPassive}

\subsubsection{Platform}
\begin{figure}[htb]
\centerline{\includegraphics[width=7cm]{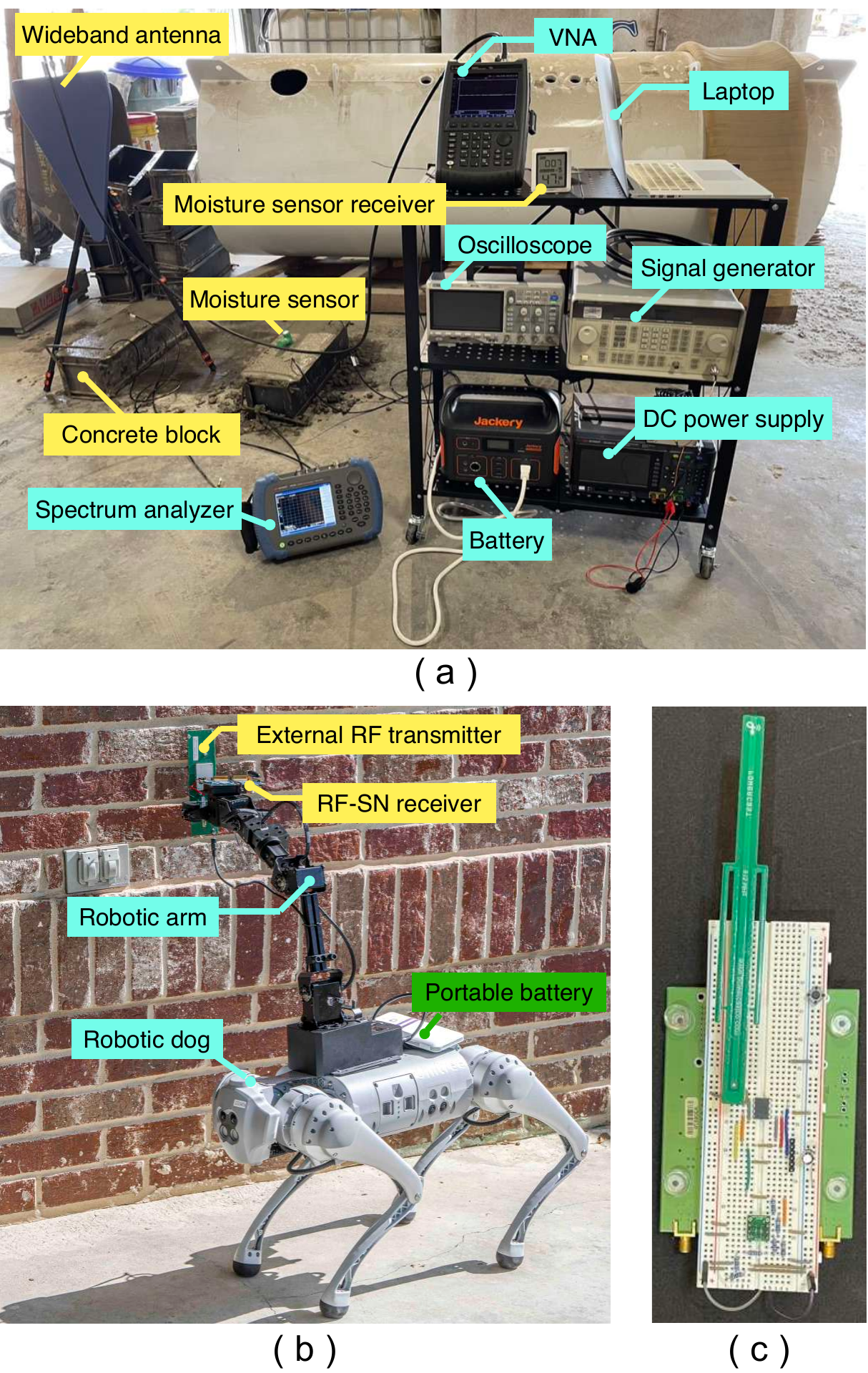}}
  \caption{Experimental platform of SHM system. (a) Experiment setup. (b) Application scenario of RF-SN with a robotic dog. (c) Prototype of the passive RF-SN.}\label{fig:AppExp}
\end{figure}

The placement of the RF transmitter was precisely controlled by a Unitree K1 robotic arm. The passive RF-SN prototype is visible in Fig.~\!\ref{fig:AppExp}\,(c). Importantly, in an effort to economize on experimental costs, only the receiving antenna was embedded within the concrete, not the entire board. To guarantee that signal reception and reflection occurred exclusively via the antenna, the board was sheathed in a tin foil material for electromagnetic interference protection. Concurrently, a coaxial cable facilitated the connection between the antenna and the board.

To receive the backscattering signal from the passive RF-SN, a 915\,MHz dipole antenna was placed at the back of the radio transmitter. The incoming signal was then enhanced by an amplifier with a 34\,dB gain before it was directed to a passive envelope detector furnished with a DC blocker. Afterward, a digital oscilloscope captured the received waveform, which was then transferred to a laptop for demodulation and symbol detection. In future iterations, the receiver could be upgraded to a universal software radio peripheral (USRP), which offers higher receiving sensitivity to further optimize signal detection performance.

It is worth noting that for a thorough evaluation of communication performance, we need to adjust the RF transmitter power across a broad range, producing carrier signals of varying intensities. However, there may be instances where the incident energy is insufficient to power the node, particularly when the oscillation frequency of the MCU is high. In such scenarios, the power source for the RF-SN node would be transitioned to a battery, ensuring continuous data transmission. In real-world applications, this situation will not happen as the node is typically configured to operate at the 4.1\,kHz signal bandwidth, which has ultra-low power consumption.

\subsubsection{BER at Various Depths}
\begin{figure}[htb]
\centerline{\includegraphics[width=6cm]{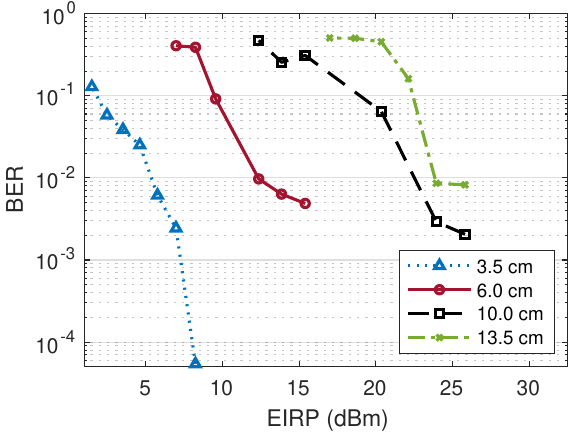}}
  \caption{BER of passive RF-SN at various concrete depths with the frequency bandwidth, symbol duration, and SF being 4.1\,kHz, 31\,ms, and 7, respectively.}\label{fig:BER_4k}
\end{figure}

Fig.~\!\ref{fig:BER_4k} compares the BER of backscattering signals received from passive RF-SNs embedded at various depths within concrete blocks 15 days after pouring. The SF value, symbol duration, and signal bandwidth are set to 7, 31\,ms, and 4.1\,kHz, respectively. With these settings, the data rate of the passive RF-SN stands at 224\,bps. Each data point is an average of 2,500 symbols. The figure demonstrates that increasing the EIRP of the RF transmitter can significantly reduce the BER across all curves. For instance, with a depth of 13.5\,cm, the BER reaches as high as 16.2\% when the EIRP is at 22.1\,dBm. This rate, however, can be reduced to 0.86\% if the EIRP is increased to 24\,dBm.

Moreover, as observed from Fig.~\!\ref{fig:BER_4k}, to maintain the same BER, the EIRP needs an approximate increment of 9\,dB for every additional 5\,cm depth into the concrete. For instance, when the node is situated 3.5\,cm deep into the concrete, the BER is 0.6\% with an EIRP of 5.8\,dBm. However, if the placement depth of the node is increased to 13.5\,cm, the EIRP should be elevated by 18.1\,dBm to preserve similar BER.

\subsubsection{BER with Different Bandwidths}
To examine the performance of the passive RF-SN across various signal bandwidths, let $r_d$, $B_w$, and $D_s$ represent the data rate in bps, signal bandwidth in Hz, and symbol duration in seconds, respectively. The interrelation of these three parameters in the square chirp modulation is expressed as follows:
\begin{equation}
\label{eq:mkas}
  r_d = \displaystyle\frac{B_w\!\times SF}{2^{\,SF}}\quad\text{and}\quad D_s = \displaystyle\frac{SF}{r_d}.
\end{equation}
 
According to equation (\ref{eq:mkas}), using square chirps with high signal bandwidth can decrease the symbol duration, thereby boosting the data rate. However, such a strategy would significantly escalate power consumption due to the necessity of a high-frequency oscillator to drive the MCU. This is highlighted in Table~\!\ref{tab:parRel}, where $F_{soc}$ represents the clock frequency to drive the MCU, and $P_c$ denotes the power consumption of the passive RF-SN. Consequently, the RF transmitter needs to augment the radiation power, which is inefficient for monitoring the health of thick concrete structures. 

\begin{table}[htp]
\scriptsize
\centering
\caption{Specifications of Passive RF-SN}
\label{tab:parRel}
\begin{tabular}{|
>{\columncolor[HTML]{C0C0C0}}c |
>{\columncolor[HTML]{EFEFEF}}c |
>{\columncolor[HTML]{EFEFEF}}c |
>{\columncolor[HTML]{EFEFEF}}c |
>{\columncolor[HTML]{EFEFEF}}c |
>{\columncolor[HTML]{EFEFEF}}c |}
\hline
$F_{soc}$ & 32.7 kHz & 1 MHz    & 2 MHz                                                  & 4 MHz     \\ \hline
$B_w$     & 4.1 kHz  & 125 kHz  &  250 kHz & 500 kHz   \\ \hline
$D_s$     & 31 ms    & 1.03 ms  & 0.51 ms                                                & 0.26 ms   \\ \hline
$r_d$     & 224 bps  & 6.8 kbps & 13.7 kbps                                              & 27.3 kbps \\ \hline
$P_c$ at $V_{dd}$ = 1.8 V     &9.3 $\mu$W	&392 $\mu$W       &418 $\mu$W     &470 $\mu$W  				\\ \hline
$P_c$ at $V_{dd}$ = 3.0 V     &26.3 $\mu$W	&850 $\mu$W       & 934$\mu$W     &1098 $\mu$W                                                       \\ \hline
\end{tabular}
\end{table}

Let $P_b$ be the bit error probability of the linear chirp. According to equation (23) in \cite{elshabrawy2018closed}, the relationship between $P_b$ and SNR is as follows:
\begin{equation}
\label{eq:jap2}
  P_b \approx \displaystyle\frac{\mathcal{Q}\left(\sqrt{SNR\cdot2^{SF+1}}-\sqrt{1.386\, SF+1.154}\right)}{2},
\end{equation}
where $\mathcal{Q}$ is the tail distribution function of the standard normal distribution.
 
With square chirps, the main lobe of the spectrum at a given time encompasses 35.6\% of the overall signal power. As a result, when the receiver uses a linear down-chirp for demodulation, the percentage of power from both the square chirp and the linear down-chirp, which contributes to signal detection, amounts to 71.2\%. Therefore, the SNR in (\ref{eq:jap2}) can be written as follows:
\begin{equation}
\label{eq:ia20}
  SNR=\displaystyle\frac{0.712\, P_s}{B_w\, N_0},
\end{equation}
where $P_s$ signifies the power of the square chirp after being enhanced by the receiving amplifier, while $N_0$ is the noise spectral density.

\begin{figure}[htb]
\centerline{\includegraphics[width=8.0cm]{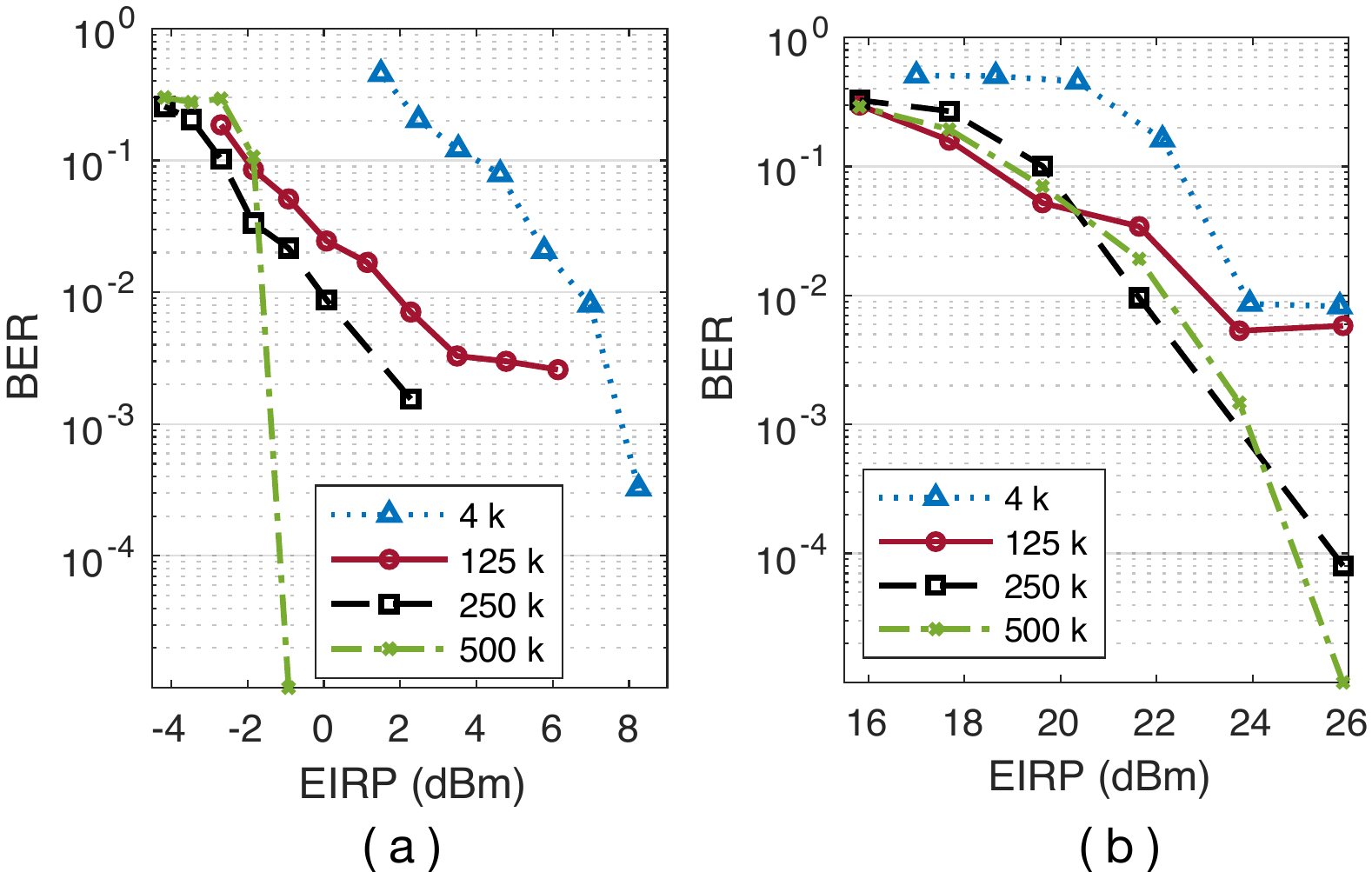}}
  \caption{BER of passive RF-SN with various bandwidth, where $\text{SF $\!=\!$ 7}$. (a)  Embedded depth $\!=\!$ 3.5\,cm. (b)  Embedded depth $\!=\!$ 13.5\,cm}\label{fig:BERDepth}
\end{figure}

Substituting (\ref{eq:ia20}) into (\ref{eq:jap2}), it can be realized that the BER of the square chirp is proportional to the signal bandwidth. However, this theoretical analysis seems to contradict the experimental findings. More specifically, in Fig.~\!\ref{fig:BERDepth}, we present the BER measured in the experiment for the passive RF-SN operating under different signal bandwidths and depths. In the figure, the data points for a bandwidth of 4.1\,kHz are an average of 2,500 symbols, while for border bandwidths of 125\,kHz, 250\,kHz, and 500\,kHz, each data point is an average of 5,070 symbols. 

Fig.~\!\ref{fig:BERDepth}\,(a) demonstrates that, given the EIRP of the radio transmitter, the passive RF-SN having a higher bandwidth can attain a lower BER. For example, when the node is situated 3.5\,cm deep within the concrete block and the EIRP is 2.5\,dBm, the BER for square chips with a 4.1\,kHz bandwidth can exceed 20.3\%. However, this value dramatically decreases to 0.71\% and 0.15\% when the signal bandwidth is expanded to 125\,kHz and 250\,kHz, respectively. 

Similar observations can be made from Fig.~\!\ref{fig:BERDepth}\,(b), where the node is embedded deeper into the concrete. As the propagation attenuation increases, the backscattering signal becomes weak. Consequently, to maintain a BER comparable to that in Fig.~\!\ref{fig:BERDepth}\,(a), the EIRP needs to be increased by at least 20\,dB. Furthermore, the BER for square chirps with a higher bandwidth continues to outperform those with a narrower bandwidth. For instance, when the EIRP of the transmitter is 22\,dBm, the BER for the signal with a bandwidth of 4.1\,kHz stands at 16.2\%. However, if the bandwidth is expanded to 500\,kHz, the BER can be reduced to 1.9\%, which is notably lower than the former scenario.

\begin{figure}[htb]
\centerline{\includegraphics[width=8.5cm]{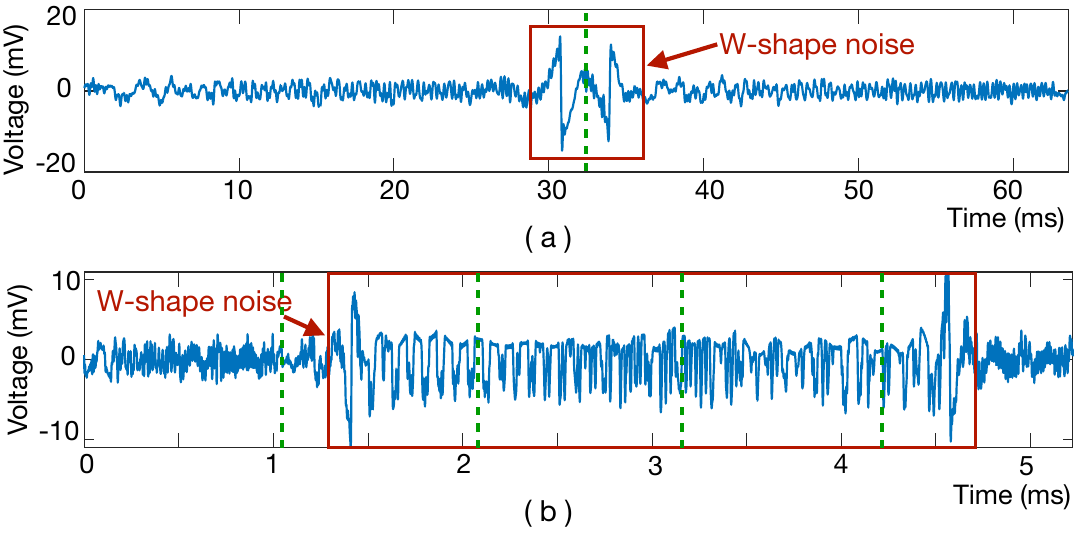}}
  \caption{Square chirp affected by W-shape interference. (a) The waveform of two continuous 'symbol 0' with 4.1\,kHz bandwidth. (c) The waveform of five continuous 'symbol 0' with 125\,kHz bandwidth.}\label{fig:NoiseImp}
\end{figure}

In order to understand the discrepancy between theory and experimental results, we carefully analyze the envelope of the received backscattering signal and discover that the W-shaped interference introduced by the RF receiver breaches the assumption of additive white Gaussian noise (AWGN) inherent in the theoretical analysis. To illustrate this observation, we select the typical 'symbol 0' influenced by the W-shaped interference and display the waveform in Fig.~\!\ref{fig:NoiseImp}. The signal frames that encompass the W-shaped interference are marked by red boxes.

As illustrated in Fig.~\!\ref{fig:NoiseImp}, each instance of W-shaped interference lasts approximately 3\,ms, and its occurrence is random, averaging around twice per second. This strong interference often leads to misclassification of data symbols, thereby causing bit errors. To simplify our analysis, we assume that in a low-SNR scenario, a symbol will be decoded incorrectly if any part of it is affected by this interference. Consequently, the symbol error rate induced by the W-shaped interference, which is denoted by $E_s$, can be roughly estimated as follows:
\begin{equation}
\label{eq:oas5}
E_s = \displaystyle \left \lceil \frac{D_w}{D_s} \right \rceil \displaystyle\frac{D_s}{T_w},
\end{equation}
where $\lceil\cdot\rceil$ refers to the ceiling function; $D_w\!=\!$ 3\,ms represents the duration of the W-shape interference, while $T_w\!=\!$ 500\,ms is the average time interval between two W-shape interferences.
 
According to (\ref{eq:oas5}), the symbol error rate generally escalates as $D_s$ increases. For instance, when the signal bandwidth is 4.1\,kHz, the symbol duration is $D_s\!=\!$ 31\,ms. In such a scenario, $E_s$ can surge to $6.2 \!\times\! 10^{-2}$. However, if the bandwidth is broadened to 125\,kHz, which in turn shortens the symbol duration to 1.03\,ms, $E_s$ then drops to $6 \times 10^{-3}$. This represents a 90\% reduction compared to the prior situation. Hence, the BER of our passive RF-SN can be improved by eliminating the W-shaped noise. This enhancement is achievable by employing a commercial USRP integrated with a low-noise amplifier for demodulation and signal detection.

\subsection{Communication Performance of Active RF-SN}
\label{subsec:ComActive}
\begin{figure}[htb]
\centerline{\includegraphics[width=5.5cm]{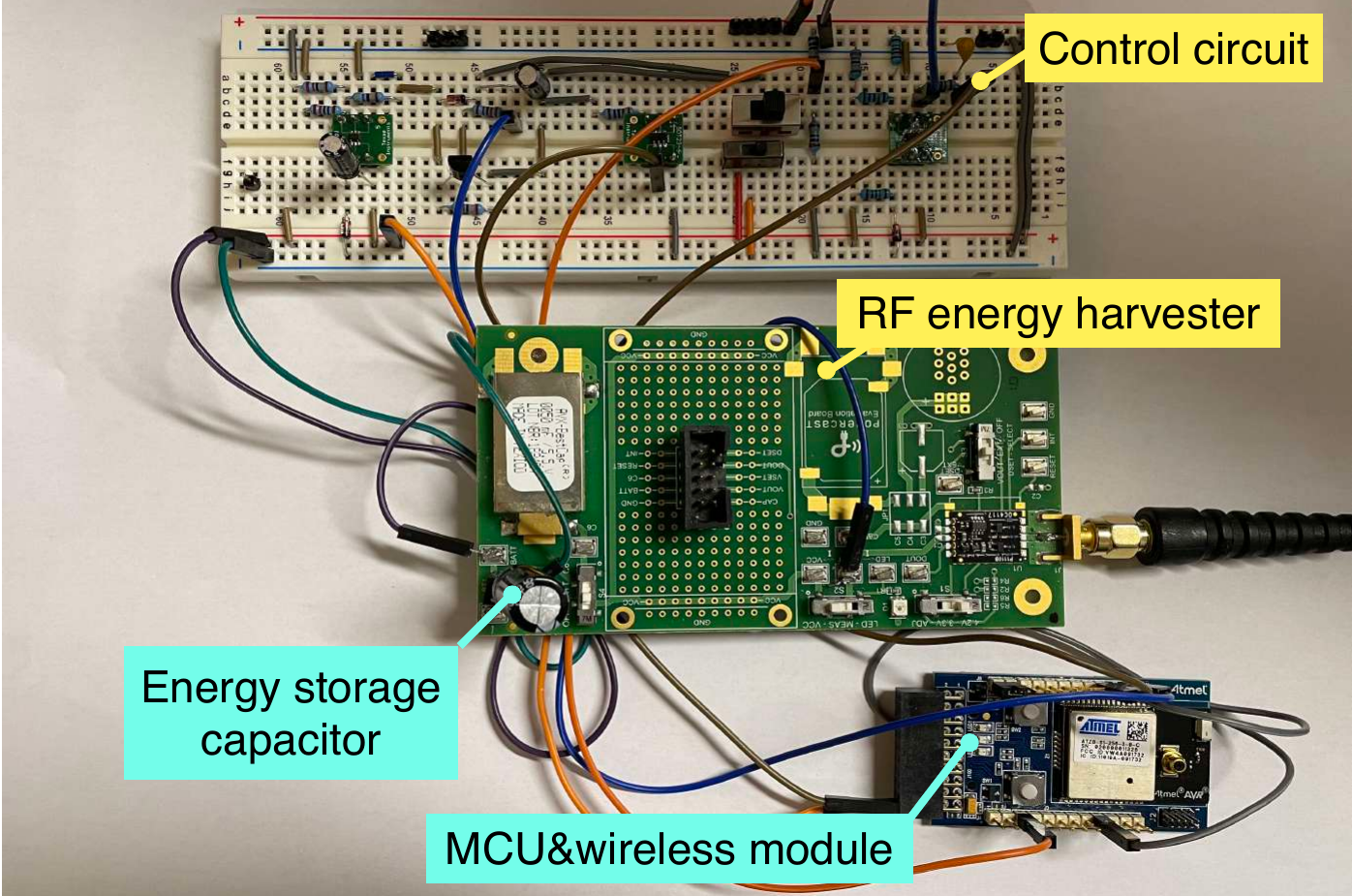}}
  \caption{Prototype of the active RF-SN.}\label{fig:ActiveRF}
\end{figure}

For active RF-SNs, the 2.4\,GHz ZigBee module integrated into the ATmega256RFR2 MCU is utilized for communication. With a receiving sensitivity of $-$100 dBm, the ATmega256RFR2 empowers the active RF-SN to transmit data over a much wider range compared to the passive one. The prototype of the active RF-SN is illustrated in Fig.~\!\ref{fig:ActiveRF}.

To evaluate the communication performance of the active RF-SN, we integrated a 2.4\,GHz dipole antenna with a directional gain of 1\,dBi into concrete blocks at varying depths. This antenna was then connected to the active RF-SN via a coaxial cable, acting as the receiver. On the transmitting end, an ATmega256RFR2 Xplained Pro evaluation kit was used to send data through a Johanson 2.4\,GHz ceramic chip antenna~\cite{microchip2015atmega256}.

In our preliminary experiment, we found that the link quality indicator (LQI) measured by the receiver reached its maximum value, which is 255, even when the transmitter was sending data at the lowest power level of $-$16.5\,dBm allowed by ATmega256RFR2. To assess the lower bounds of communication performance for the active RF-SN, we decided to attenuate the transmitter's radiation power. This was done by housing the transmitter within a sealed metal box and calibrating the received power inside a microwave anechoic chamber, thereby mitigating multipath effects and interference from the surrounding environment.

\begin{figure}[htb]
\centerline{\includegraphics[width=8.5cm]{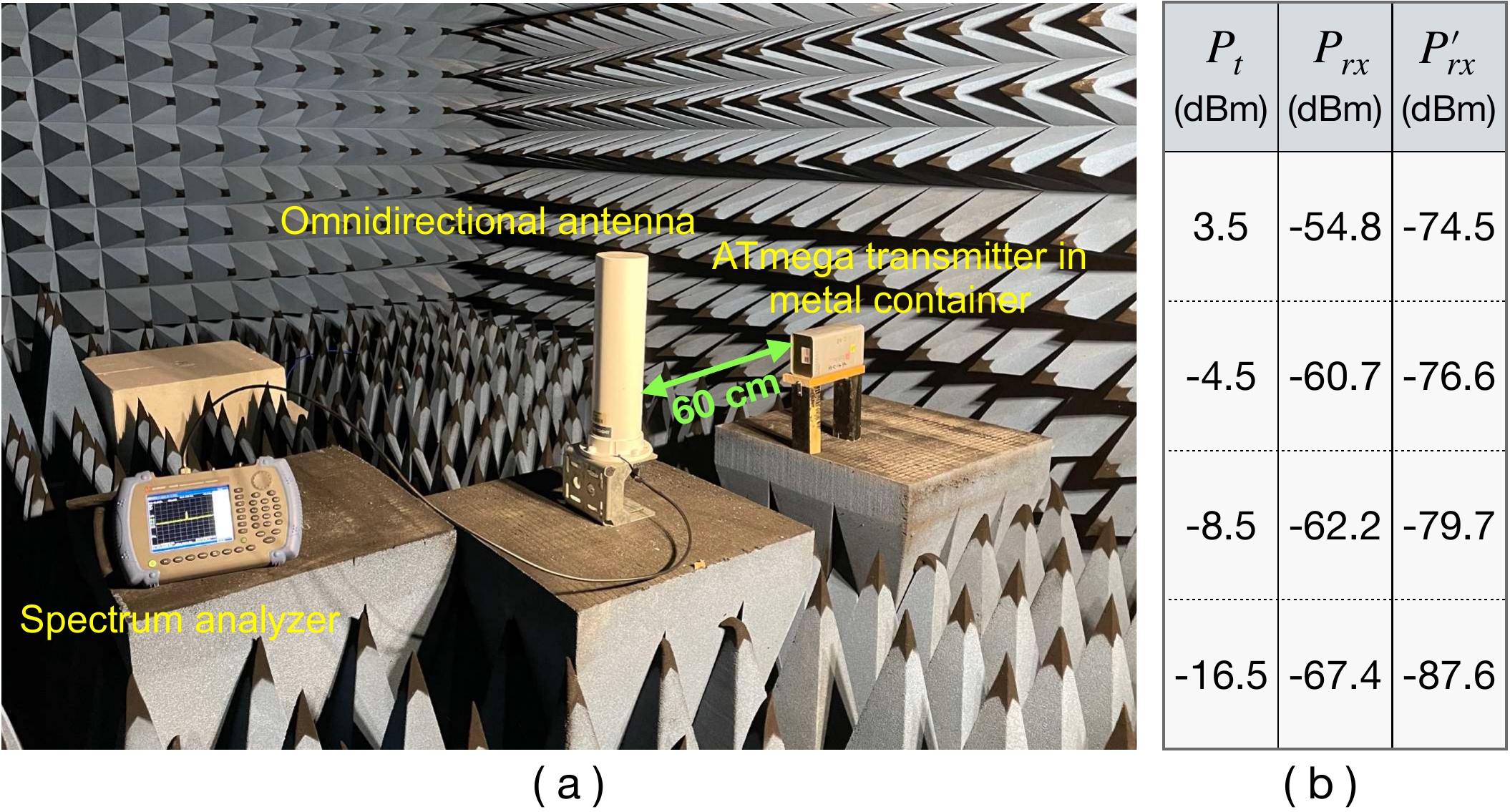}}
  \caption{Calibration of radiation power in the chamber. (a) Experimental setup. (b) Receiving power with and without metal box.}\label{fig:chamber}
\end{figure}

To calibrate the radiation power of ATmega256RFR2, we use a Keysight N6850A omnidirectional antenna with 0 dBi gain at 2.4\,GHz \cite{keysight2015n6850A} to receive the RF signal from the transmitter, as shown in Fig.~\!\ref{fig:chamber}\,(a). The distance between the transmitter and the receiver is 60\,cm, and the signal strength is measured using the Agilent N9340B spectrum analyzer. We denote the original transmission power of ATmega256RFR2 as $P_t$. Additionally, let $P_{rx}$ and $P^{\prime}_{rx}$ represent the power received by the N6850A antenna when the transmitter is exposed in the open air and enclosed in a metal box, respectively. The relationships between $P_t$, $P_{rx}$, and $P^{\prime}_{rx}$ measured through the experiment are presented in Fig.~\!\ref{fig:chamber}\,(b).

\begin{figure}[htb]
\centerline{\includegraphics[width=8.0cm]{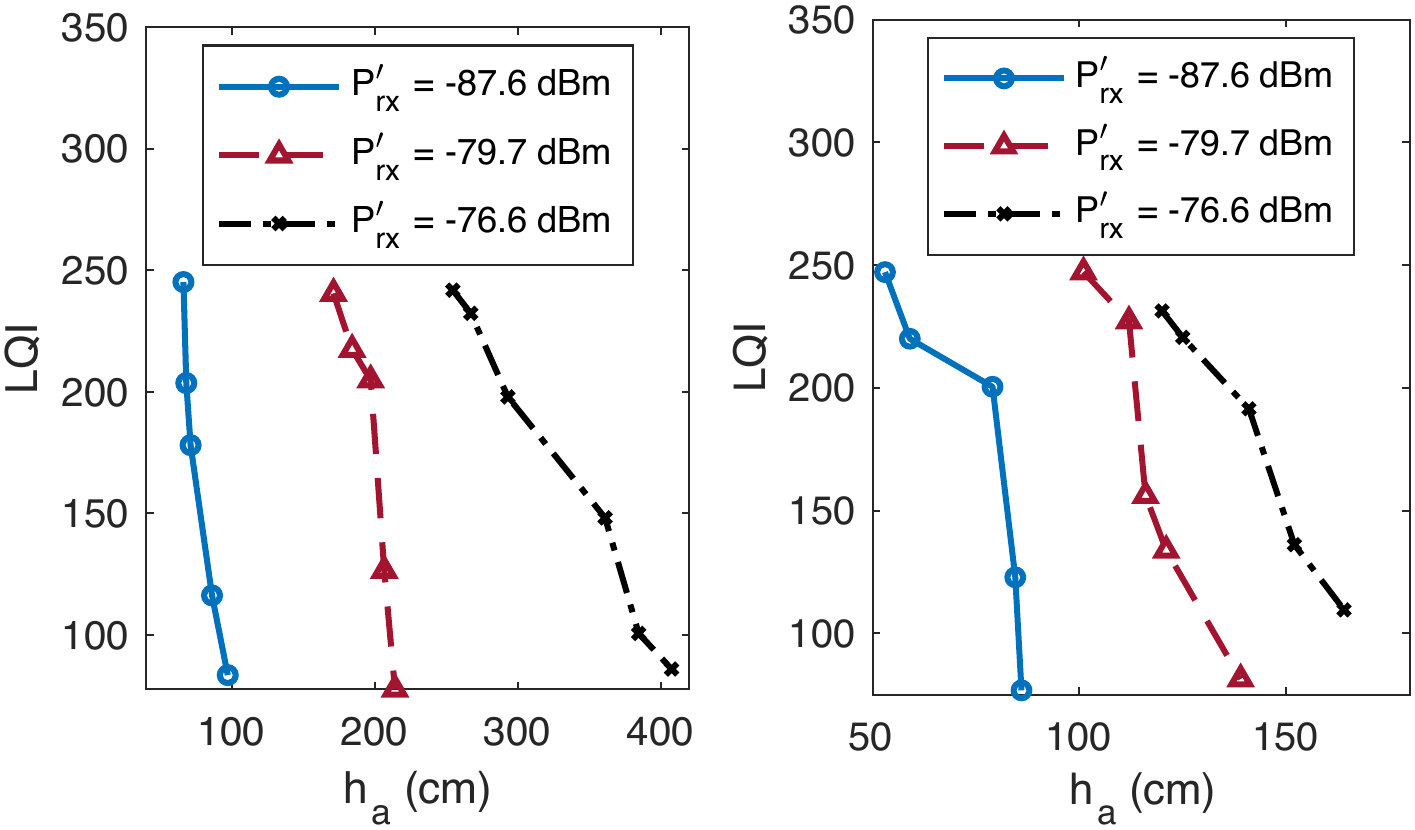}}
  \caption{LQI of active RF-SN. (a) Embedded depth $\!=\!$ 3.5\,cm. (b)  Embedded depth $\!=\!$ 13.5\,cm}\label{fig:LQI}
\end{figure}

Based on the transmission power calibrated in Fig.~\!\ref{fig:chamber}\,(b), we depict the LQI measured by the active RF-SN embedded at different depths within the concrete block in Fig.~\!\ref{fig:LQI}, where the data rate is 250\,kbps and $h_a$ is the distance between the transmitter and the surface of the concrete block. 

As demonstrated in Fig.~\!\ref{fig:LQI}, the active RF-SN leveraging the ZigBee protocol can accomplish a relatively long communication range even if the radiation power is low. Fig.~\!\ref{fig:LQI}\,(a) indicates that for a sensor node located 3.5\,cm within the concrete, the LQI can achieve a value of 255 when $h_a$ is 66\,cm and $P^{\prime}_{rx}$ is only $-$87.6\,dBm. However, $h_a$ can be extended to as much as 255\,cm, given an increase of $P^{\prime}_{rx}$ to $-$76.6\,dBm.

A comparative analysis of Fig.~\!\ref{fig:LQI}\,(b) and Fig.~\!\ref{fig:LQI}\,(a) reveals that, given $P^{\prime}_{rx}$, the communication range significantly diminishes if the burial depth of the RF-SN increases due to the substantial attenuation of the 2.4\,GHz signal within the concrete. For instance, at $P^{\prime}_{rx}$ of $-$79.7\,dBm, an LQI of 125 can be achieved with an $h_a$ of 206\,cm. In this situation, the packet error rate is 50\%. However, for the same LQI and $P^{\prime}_{rx}$ value, increasing the embedding depth to 13.5\,cm reduces $h_a$ to just 130\,cm, significantly shorter than the prior instance. This underscores the impact of RF-SN burial depth on communication range in such settings.

\section{Conclusion}
\label{sec:Con}
In this study, we delved into the potential of radio energy-powered sensor nodes for long-term and mobile structural health monitoring. We developed both active and passive RF-SNs to accommodate different application needs. According to our experimental results, an active RF-SN embedded at 13.5\,cm depth within concrete can be effectively powered by a 915\,MHz radio transmitter with an EIRP of 32.5\,dBm, enabling sensing and reliable communication via the ZigBee protocol. Simultaneously, a passive RF-SN situated at the same depth within the concrete allows for a decrease in the transmitter's EIRP to 23.6\,dBm. In this scenario, the passive RF-SN can continually transmit data at a speed of 224\,bps using square chirp-based backscattering modulation. The RF-SN presented in this paper offers a promising pathway toward battery-free, enduring, and automated structural health monitoring.

\section*{Acknowledgment}
\label{sec:Ack}
The authors thank MMC Materials, Inc. in Starkville, MS and Mr. Gary Scott for providing all the support for the experiment. This work is supported in part by the US National Science Foundation under Awards CIF-2051356, CNS-2122167, and CNS-2122159.

\bibliographystyle{IEEEtran}
\bibliography{Concrete}

\end{document}